\documentclass[%
reprint,
superscriptaddress,
 nofootinbib,
 amsmath,amssymb,
 aps,
prb,
floatfix,
 showkeys,
]{revtex4-2}

\usepackage{graphicx}
\usepackage{dcolumn}
\usepackage{bm}
\usepackage{siunitx}
\usepackage{makecell}
\usepackage{collcell}
\usepackage[version=4]{mhchem}
\usepackage{hyperref}
\usepackage{textcomp}
\usepackage{etoolbox}

\usepackage{cleveref}
\crefname{section}{sec.}{secs.}

\usepackage[caption=false]{subfig}
\newcommand{\phantomsubfloat}[1]{
    {
        \captionsetup[subfigure]{labelformat=empty}
        \subfloat[][]{#1}
        \vspace{-5pt}
    }%
}

\usepackage{comment}
\usepackage{multirow}
\usepackage{orcidlink}
\usepackage{amsmath}
\usepackage{enumerate}
\usepackage{titlesec}

\newcolumntype{P}{
    >{\collectcell}r<{\endcollectcell}
		@{${}\propto{}$}
	>{\collectcell}l<{\endcollectcell}}

\newcommand{\abs}[1]{
   \left|#1\right|
}

\usepackage{chngcntr}
\counterwithin*{section}{part}
\counterwithin*{table}{part}
\counterwithin*{figure}{part}

\usepackage{titlesec}

\titleformat{\part}[display]
  {\normalfont\centering\bfseries}
  {}
  {1pc}
  {\MakeUppercase}

\begin{document}

\title{The domain-wall/metal-electrode injection barrier in lithium niobate: \texorpdfstring{\\}{} Which electrical transport model fits best?}

\author{Manuel~Zahn\,\orcidlink{0000-0003-0739-3049}}
\affiliation{Institute of Applied Physics, Technische Universit\"at Dresden, 01062 Dresden, Germany}
\affiliation{Experimental Physics V, Center for Electronic Correlations and Magnetism, University of Augsburg, 86159~Augsburg, Germany}

\author{Elke~Beyreuther\,\orcidlink{0000-0003-1899-603X}}%
 \email{elke.beyreuther@tu-dresden.de}
\affiliation{Institute of Applied Physics, Technische Universit\"at Dresden, 01062 Dresden, Germany}

\author{Iuliia~Kiseleva\,\orcidlink{0009-0002-5435-056X}}%
\affiliation{Institute of Applied Physics, Technische Universit\"at Dresden, 01062 Dresden, Germany}

\author{Julius~Ratzenberger\, \orcidlink{0000-0001-6896-2554}}
\affiliation{Institute of Applied Physics, Technische Universit\"at Dresden, 01062 Dresden, Germany}
\affiliation{Würzburg-Dresden Cluster of Excellence (EXC~2147) ctd.qmat -- Complexity, Topology, and
Dynamics in Quantum Matter, 01062 Dresden, Germany}

\author{Michael~R\"using\,\orcidlink{0000-0003-4682-4577}}
    \affiliation{Institute for Photonic Quantum Systems (PhoQS), Department of Physics, University of Paderborn, 33098 Paderborn, Germany}

\author{Lukas~M.~Eng\,\orcidlink{0000-0002-2484-4158}}
\affiliation{Institute of Applied Physics, Technische Universit\"at Dresden, 01062 Dresden, Germany}
\affiliation{Würzburg-Dresden Cluster of Excellence (EXC~2147) ctd.qmat -- Complexity, Topology, and
Dynamics in Quantum Matter, 01062 Dresden, Germany}

\date{\today}

\begin{abstract}    
The comprehensive description of both the electrical transport along conductive domain walls (CDWs) in lithium niobate (LNO) single crystals and the charge injection at the interfacing metal electrodes, emerged to be a complex challenge. Recently, a heuristic evaluation allowed to postulate the "$R2D2$" equivalent-circuit model (consisting of two parallel resistor-diode pairs) to appropriately match the DC current-voltage (I-V) characteristics. Here, we carefully revisit the interfacial electrical behavior, i.e., the diode part of the equivalent circuit model, since many more processes beyond the diode-related electron hopping transport (HT) assumed so far, may concurrently occur, such as thermionic emission (TE), Fowler-Nordheim tunneling (FNT), space-charge limited conduction (SCLC), and others more. The "$R2D2$" model thus needs to be generalized into an "$R2X2$" circuit model (with X = HT, TE, FNT, and others) to fit to the experimental data. Moreover, to double check for the best I-V curve fitting to the different theories, we apply a higher-harmonic DW current-contribution (HHCC) analysis, i.e., an AC I-V inspection, that allows us to discriminate between all these possible models with much higher precision than from pure DC I-V curve fitting. Both the AC and DC analysis reveal well consistent results, finally finding that the FNT model accounts best for the domain-wall/electrode junctions investigated here.
\end{abstract}


\maketitle

\section{Introduction}
\label{sec:intro}
    
Ferroelectrics have been proposed as versatile functional materials to overcome limits of conventional semiconductor-based computing architectures, both within classical von-Neumann architectures via field-effect transistors and also completely new concepts such as reservoir computing \cite{meier_ferroelectric_2022, mikolajick_ferroelectric_2023, everschor-sitte_topological_2024}. In particular, \emph{confined} electronic transport along ferroelectric \emph{charged domain walls}, which allows for 2-dimensional current flow, has been subject of intense research during the last decade \cite{catalan_domain_2012,meier_functional_2015,sluka_charged_2016,bednyakov_physics_2018,sharma_functional_2019,nataf_domain-wall_2020,meier_ferroelectric_2021,sharma_roadmap_2022}. Concrete implementations of such low-dimensional ferroelectric components in diodes \cite{schaab_electrical_2018,Niu_Diode_2023}, memory devices \cite{Jiang_Ferroelectric_2020,Hu_Cryogenic_2024,Yu_Adjustable_2026}, in field-effect transistors \cite{chai_nonvolatile_2020}, or for matrix multiplication via memristors \cite{seufert_crossbar_2021} underline these efforts.

To fully unravel the potential of ferroelectric domain-wall (nano-)electronics, the analysis of the electrical transport properties of both the domain walls themselves as well as of their \emph{interfaces} with the (typically metallic) contact electrodes became of raising interest and motivated a systematic survey on conduction mechanisms, which turned out to be a rather complex task. For each ferroelectric host material and its specific domain wall (DW) type(s) as well as for each combination with a given electrode metal, a separate investigation appears to be fundamentally required, because different mechanisms -- most of them known from solid-state and semiconductor physics -- have been reported in conjunction with ferroelectrics in the past (see \cref{tab:conduction_mechanisms}).

\begin{table*}
	{\renewcommand{\arraystretch}{1.3}
    \begin{tabular}{llrlr} \hline \hline
		\thead{Mechanism} & \thead{Abbreviation} &
		      \multicolumn{2}{c}{\thead{I-V characteristic}} &
		      \thead{FE example} \\ \hline
		Ohmic transport & OT & \(I=\) & \(\sigma_\text{OT} \, U\)
            & \ce{BiFeO3} (bulk) \cite{seidel_conduction_2009} \\
		Hopping transport & HT & \(I=\) & \(I_\text{HT} \left(\exp\left(U/U_\text{HT}\right) - 1\right)\)
            & \ce{LiNbO3} (bulk) \cite{werner_large_2017} \\
		Space charged limited conduction & SCLC & \(I=\) &
            \(\alpha_\text{SCLC} \, U^2\) 
            & \ce{GaV4S8} (bulk) \cite{puntigam_strain_2022} \\ 
		Poole Frenkel emission & PFE & \(I =\) & \(\sigma_\text{PFE} \, U \exp\left(
            \sqrt{U / U_\text{PFE}}\, \right)\)
            & \ce{Er(Mn,Ti)O3} (bulk) \cite{holstad_electronic_2018} \\ \hline
		Thermionic emission & TE & \(I=\) &
            \(I_\text{TE} \, \exp\left(\sqrt{U/U_\text{TE}}\right)\)
            & \ce{BaTiO3} (NS) \cite{qi_modified_2015} \\
        Thermionic-field emission & TFE & \(I=\) &
            \(\sigma_\text{TFE} \, U \exp \left(U^2 / U_\text{TFE}^2 \right)\)
            & \ce{SrBi2Ta2O9} (bulk) \cite{nagasawa_imprint_1999} \\
		Fowler-Nordheim tunneling & FNT & \(I=\) & \(\alpha_\text{FNT} \, U^2 \exp\left(U_\text{FNT} / U\right)\)
            & \ce{PbZr_{0.2}Ti_{0.8}O3} (NS) \cite{garcia_ferroelectric_2014} \\ \hline \hline
	\end{tabular}}
	\caption{Overview on electric conduction mechanisms, which were observed in bulk materials and nanostructures (NS) of conductive ferroelectrics (cf. review article by \citet{chiu_review_2014}) including the specific shape of the current-voltage (I-V) curve and corresponding literature examples. The I-V characteristics describe the behavior of the electric voltage \(U\) as a function of the current \(I\) and include model-specific fit parameters (in units of a current $I_\text{\dots}$, a voltage $U_\text{\dots}$, a conductivity $\sigma_\text{\dots}$, or more specific: $\alpha_\text{\dots}$). The mechanisms listed below the horizontal line are commonly known as \emph{interface-limited} processes, while the ones above are called \emph{bulk-limited}.}
	\label{tab:conduction_mechanisms}
\end{table*}

For the model system of hexagonal ferroelectric domain walls with enhanced conductivity in lithium niobate (\ce{LiNbO3}, LNO) single crystals, which will be the focus of this work, the struggle for a comprehensive understanding of the underlying electrical transport mechanisms is particularly well-documented over the last decade (e.g. refs.~ \cite{schroder_conducting_2012,schroder_conductive_2014,werner_large_2017,ratzenberger_reproducible_2024,zahn_equivalentcircuit_2024}). In principle, LNO's uniaxial ferroelectricity enables a deterministic domain-wall-geometry manipulation \cite{kirbus_realtime_2019} and makes it a prospective candidate for domain-wall implemented electronic circuits. After substantial DC conductivity in LNO DWs under UV-light \cite{schroder_conducting_2012} and in the dark \cite{werner_large_2017} had been shown successfully within several pioneering works, decisive improvements on the conductive-domain-wall preparation have been made by finding a protocol to more reproducibly enhance the DW conductivity by high-voltage ramping \cite{godau_enhancing_2017, ratzenberger_reproducible_2024} to create long-term conductive DWs with predictable I-V~characteristics. However, there is still a remarkable variety of I-V characteristics observed despite of using identical process parameters for CDW preparation, interestingly, it became more and more clear that the real structure of the \emph{interfaces} between ferroelectric DWs and the metallic contact electrodes in general, and the LNO-DW/Cr-electrode-system in particular, play a key role for the electrical behavior of the whole system. This has been taken into account within a recent previous work \cite{zahn_equivalentcircuit_2024}, where extensive I-V~measurements have been conducted and evaluated, an equivalent circuit model has been proposed that consists of a parallel connection of two resistor/diode-pairs (the "\textit{R2D2} model"), and a thermally activated hopping process has been inferred from the I-V~curves' temperature dependence, while various other hopping processes such as variable-range hopping could be excluded. However, although the assignment of (i) the resistors of the \textit{R2D2} model to the transport along the DWs and (ii) the diode-like part to the transport across the electrode/DW interface looks plausible at a first glance, the mathematical description of the interface transport by the well-known Shockley diode equation might not be the best-fitting solution, especially not in the view of the large number of alternative processes as listed in \cref{tab:conduction_mechanisms} and due to the fact that the search for the best-fitting interface transport model has shown to be a very elaborate task also for other ferroelectric conducting domain walls in the past \cite{Guyonnet_Conduction_2011}.

The present work addresses this open issue using the following two refined investigative strategies: 

First, the I-V~characteristics of two exemplary lithium niobate domain wall samples with enhanced conductivity will be \emph{re-evaluated} by fitting the diodic part of the \textit{R2D2} model additionally with alternative (interface-)transport models that have been proposed to account for electrical conduction in other ferroelectrics [space-charge limited conduction (SLC), thermionic emission (TE), Fowler-Nordheim tunneling (FNT)] or that can generally occur at oxide/metal interfaces [thermionic field emission (TFE)], resulting in a generalized "\textit{R2X2}" model, and the fit residuals will be assessed. 

Second, an experimental scenario with much higher sensitivity as compared to standard DC I-V~characterization will be introduced, namely a higher-harmonic current contribution (HHCC) analysis upon alternating-voltage excitation, in order to validate the conclusions drawn from the pure I-V~curve fitting procedure.

\begin{figure}[htb]
    \centering
    {\phantomsubfloat{\label{fig:lno_dc_dw1}}
     \phantomsubfloat{\label{fig:lno_dc_dw2}}}
    \includegraphics[width=\linewidth]{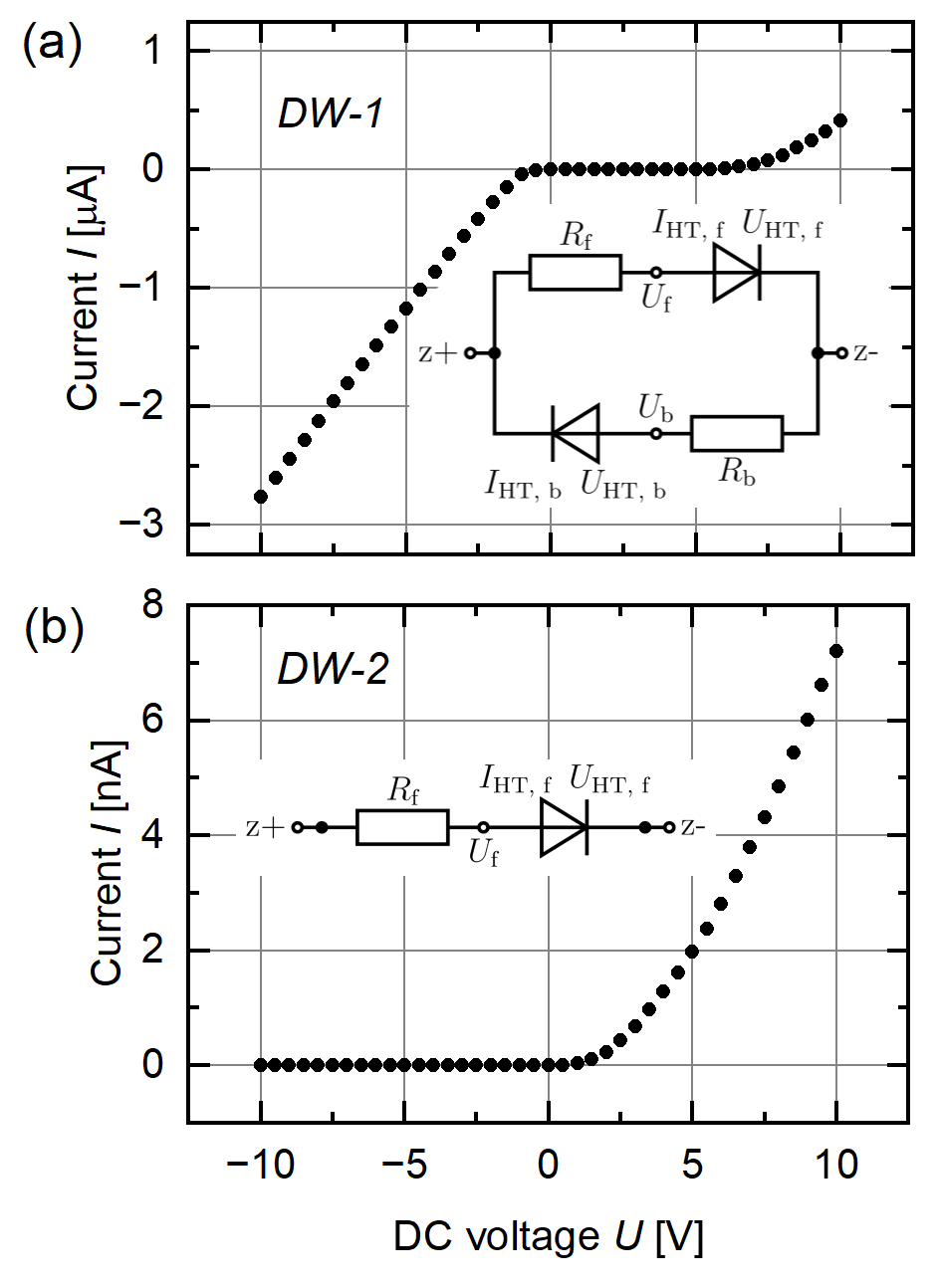}
    \caption{DC I-V characteristics of the two conductive \ce{LiNbO3} domain-wall samples of the present study and their equivalent circuits. \textbf{(a)} I-V curve of sample \textit{DW-1} that can be fitted with the "\textit{R2D2}" model (inset) of ref.~\cite{zahn_equivalentcircuit_2024}, which would correspond to an assignment of the forward and reverse diode-like circuit parts to hopping transport (HT) in terms of \cref{tab:conduction_mechanisms}. \textbf{(b)} In contrast, the reverse current path is only weakly developed in the second exemplary sample of this work, \textit{DW-2}, exhibiting a rectifying behavior, where the "\textit{R2D2}" circuit model can be reduced to \textit{one} branch: "\textit{RD}". A "positive" voltage means here that the positive electrode is connected to the z+ side of the LNO crystal. The indices $f$ and $b$ refer to the forward and backward direction, while $R$, $I_\text{HT}$, and $U_\text{HT}$ symbolize the resistance, the diode's saturation current, and the diode's characteristic voltage (which bears the ideality factor), respectively (also cf.~ref.~\cite{zahn_equivalentcircuit_2024}).}
    \label{fig:lno_iu_curves}
\end{figure}

\section{Materials and methods}

\subsection{\texorpdfstring{Preparation of conductive \ce{LiNbO3}}{LiNbO3} domain walls, acquisition and fitting of current-voltage curves} 

\subsubsection{Domain growth by UV-assisted liquid-electrode poling}

Initially, two samples were cut from a monodomain, \SI{5}{\mol\percent\ \ce{MgO}}-doped, congruent, 200-\textmu m-thick, z-cut \ce{LiNbO3} wafer by \textit{Yamaju Ceramics Co., Ltd.}. The two samples, labeled \textit{DW-1} and \textit{DW-2} in the following, measure \(5 \times \SI{6}{\milli\meter\squared}\) along their crystallographic x- and y-axis, respectively. Realizing the protocols described in detail earlier \cite{godau_enhancing_2017,godau_herstellung_2018,ratzenberger_reproducible_2024}, one single hexagonally-shaped reversely polarized domain (diameter approx.~100~\textmu m) was grown by UV-laser-assisted poling into each sample. In other words, domain walls that touch both surfaces, having, strictly mathematically speaking, two boundary components and genus one, were created. The 3D close-up view within \cref{fig:ac_principle:circuit} gives a schematic impression of the domain wall geometry.

\subsubsection{Evaporation of electrodes}

In a next preparation step, macroscopic 10-nm-thick Cr electrodes were vapor-deposited onto both crystal surfaces covering the DWs completely (cf.~\cref{fig:ac_principle:circuit}). For electrical reference measurements of the bulk properties, a second pair of such electrodes on a neighboring mono\-domain part of each sample was deposited in parallel during the same evaporation run, as also illustrated in \cref{fig:ac_principle:circuit}.
The as-grown domain walls were electrically tested by recording \SI{\pm 10}{V} standard current-voltage (I-V) characteristics that revealed a very low, nearly bulk-like conductivity with currents in the 0.1-pA-range.

\subsubsection{"Enhancement" of the DW conductivity by high-voltage ramping and final I-V characteristics}

Subsequently, the DW conductivity was \emph{enhanced} according to the protocol of ref.~\cite{godau_enhancing_2017} by linearly ramping up a high voltage of \SI{450}{V} over a time period of \SI{75}{s}, provided by the voltage source of a \textit{Keithley 6517B} electrometer. As a result, the resistance of the DWs decreased significantly by three to six orders of magnitude, as seen from the samples' final static I-V characteristics, acquired by current-voltage sweeps (\SI{0.33}{V/s}, \SI{0.5}{V} steps) with the above-mentioned electrometer (\cref{fig:lno_iu_curves}): 
\begin{itemize}
    \item \textit{DW-1} (\cref{fig:lno_dc_dw1}) shows a conductive behavior for both directions of the electric field, so both channels within the recently proposed double-resistor-double-diode (\textit{R2D2}) model (inset) \cite{zahn_equivalentcircuit_2024} are active, while
    \item for sample \textit{DW-02} (\cref{fig:lno_dc_dw2}) the backward channel is insufficiently developed, so conductance is observed only in forward direction; consequently it will be modeled as a single path with one resistor and one diode, only. 
\end{itemize}
Though in terms of reproducibility not really intended, these significantly different I-V characteristics give both samples a complementary character and the details of our later HHCC analysis can be studied in these two different circuit scenarios. Note that the temperature-dependent conduction behavior of \emph{DW-1} has already been reported earlier~\cite{zahn_equivalentcircuit_2024} and that in the present study the I-V-curve acquisition range for this sample was shifted from -10/+10~V to -5/+15~V in order to record more data points from the non-linear part of the curve, which is essential for the later HHCC analysis.

\subsubsection{I-V curve fits with different \texorpdfstring{R2X2}{\textit{R2X2}} models}
\label{sec:iv_fits}

Analogously to our previous work, which considered solely resistor-diode combinations \cite{zahn_equivalentcircuit_2024}, I-V curves for the \emph{different models} of \cref{tab:conduction_mechanisms} have been calculated by numerically evaluating the continuity equations for the current at the intermediate nodes between the resistor and the non-linear circuit elements (open circle nodes in the inset of \cref{fig:lno_dc_dw1}). This resolves the intermediate potential at these nodes ($U_\text{f}$ and $U_\text{b}$) and, subsequently, via the single elements' I-V characteristics the total current through the circuit. The underlying algorithm has been encapsulated by a least-square optimization algorithm, which varies the four to six model parameters in order to minimize the logarithmic difference $\mathcal{D}$, which will be referred to as the \emph{sum of residuals} from now on, between experimental and fitted I-V curve:
\begin{equation}
    \mathcal{D} = \sum_i {\underbrace{\left( \log_{10}\frac{\abs{I_\text{exp.}(U_i)}}{\abs{I_\text{model}(U_i)}}\right)}_{:= \mathcal{D}_i}}^2,
    \label{equ:residuals}
\end{equation}
while taking into account the current's measurement uncertainty of around \num{2.5} orders of magnitude below the maximum current and skipping data points, where the measured current values are below that. The single logarithmic differences $\mathcal{D}_i$ (with $i$ symbolizing the data point index) between the measured current $I_\text{exp.}$ and the current according to the fit curve $I_\text{model}$ at a given voltage $U_i$ will be tagged \emph{residuals} throughout the following text.

\subsection{Measurement of higher-harmonic current contributions (HHCCs)} 
\label{sec:hhcc_derivation_main}

\begin{figure*}[htb]
	\centering
	{\phantomsubfloat{\label{fig:ac_principle:circuit}}
     \phantomsubfloat{\label{fig:ac_principle:IVcurve}}
     \phantomsubfloat{\label{fig:ac_principle:complex_plane}}}
    \includegraphics[width=\textwidth]{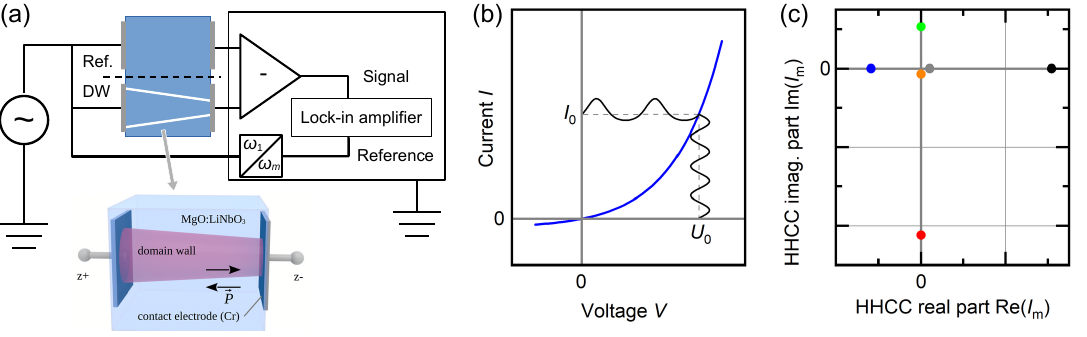}
	\caption{Principle of higher-harmonic current contributions'~(HHCC) acquisition of a structure consisting of a conductive ferroelectric domain wall in LNO single crystal contacted with Cr electrodes on the z+ and z- side. \textbf{(a)} Scheme of the electric circuit including signal generator, sample with two pairs of electrodes of the \textit{same} area (one contacting the DW, the other contacting the pure bulk as a reference), and lock-in amplifier. The sample incorporates here an artificially poled single cylindrical ferroelectric DW as shown in the 3D close-up view. More details on the experimental setup are provided in sec.~C of the Supplemental Material \cite{supplement}. \textbf{(b)} Within the HHCC measurement, a sinusoidal voltage $U(t)$ is applied around the DC offset voltage \(U_0\), see \cref{equ:exc_voltage}. Due to non-ohmic conduction behavior (here, exemplarily, the I-V curve of a single diode is shown), the electric current, as induced by the excitation field, follows a non-harmonic oscillation with the same periodicity in time as the excitation signal, which can be expressed in the trigonometric orthonormal base of sine and cosine functions with angular frequency \(\omega_1\) and their integer multiples \(\omega_m = m \cdot \omega_1\) (Fourier series). The HHCCs are characterized by their harmonic order \(m\) and their complex amplitude \(I_m\). \textbf{(c)} Nyquist diagram of the complex Fourier coefficients of the electric-current response, illustrated for the single-diode case, exhibiting a characteristic pattern in amplitude and phase as derived in \cref{sec:methods:math} and represented in \cref{equ:diode_fourier_coefficients}. The harmonic orders are color-coded from \(m = 1\) to \(m = 6\) in black, red, blue, green, grey, and orange, respectively.}
	\label{fig:ac_principle}
\end{figure*}

To probe and analyze features of the non-ohmic parts of the current-voltage characteristics that can not be resolved via static DC I-V recordings, we use the \textit{differential} conductance measurement scheme sketched in \cref{fig:ac_principle:circuit}. To get rid of possible DC bias offsets, an excitation voltage with the form of a sine function is applied on \textit{two pairs} of electrodes on the LNO crystal, one with and one without enclosing a domain wall in between, by a waveform generator:

\begin{equation}
    U(t) = U_0 + U_1 \sin(\omega_1 t). 
    \label{equ:exc_voltage}
\end{equation}

Here, $U_0$ is the offset voltage, $U_1$ the amplitude, $\omega_1$ the excitation angular frequency, and the sine function is chosen (instead of the symmetric cosine) for practical reasons, i.e., because the lock-in-amplifier based signal analysis "works" with this convention, i.e., all reference waves share the zero crossing on the rising edge with the signal data. The usage of the specific two-electrode-pair structure allows (after current-to-voltage conversion) to subtract the contribution of the capacitor formed by the electrodes from the contribution of the capacitor formed by the actual electrode/DW-structure in-situ using analog electric circuity, and thus to detect the \textit{net domain-wall current} response, which is non-harmonic but exhibits the same time periodicity as the excitation voltage $U(t)$ (illustrated in \cref{fig:ac_principle:IVcurve}). The integer Fourier components $I_m$ of this signal [in the following also termed higher-harmonic current contributions (HHCCs)], represented by separate amplitudes ($\abs{I_m}$) and phases ($\arg{I_m}$) for each harmonic order $m$ (sketched in \cref{fig:ac_principle:complex_plane}), characterizing the current contribution at the angular frequency $\omega_m = m \cdot \omega_1$, are recorded using a lock-in amplifier. Thereby the excitation sine wave acts as the reference signal, while extracting the $m^\text{th}$ harmonic order of the reference wave as discussed in the following section. Two setups have been assembled, facilitating either a "fast-acquisition" by synchronous demodulation of multiple harmonics, which was applied for the investigations of \textit{DW-1}, and a "high-precision" version for detecting very low currents, as required and applied for the higher-resistive case of \textit{DW-2}. Both setups are discussed in more detail in sec.~C of the Supplemental Material \cite{supplement}.

Here, the HHCCs were recorded from first up to sixth order ($m = 1 \dots 6$) as a function of the offset voltage $U_0$ (chosen typically between \SI{-10}{V} and \SI{+10}{V}) and the amplitude $U_1$ (\SI{1}{mV} up to \SI{3}{V}), while the frequency dependence (from \SI{10}{Hz} up to \SI{10}{kHz}) is used as a diagnostic tool, ensuring that the measurement is conducted in a "DC-like" regime, which means the absence of imaginary first order current contributions [$\arg(I_1) = 0$]. Note that the characterization and suppression of harmonic distortions by the experimental setup itself, i.e., from the signal generator, are essential, because they may generate additional HHCCs, which are hard to separate from the sample-induced signal \cite{zahn_nonlinear_2022}. To check this issue, the absence of harmonic distortions by the experimental setup itself was proven with a characterization measurement of a commercial Schottky diode that is discussed in sec.~D of the Supplemental Material \cite{supplement}. 

As the analysis of higher harmonic currents is in principle possible over a very large parameter space spanned by $U_0$, $U_1$, the number of harmonic orders $m$, and $\omega_1$, we selected several particularly interesting measurement ranges and parameters:
\begin{itemize}
    \item On sample \textit{DW-1} -- motivated by the two strongly non-linear regions of the I-V~curve around $U_\text{DC} = \SI{0}{V}$ and \SI{5}{V} -- the HHCCs were probed as a function of the offset voltage $U_0$ between -5 and +15~V, up to the sixth harmonic order ($m = 1 \dots 6$), at constant frequency ($\omega_1/2\pi=\SI{1.5}{kHz}$) and excitation amplitude ($U_1 = \SI{0.71}{V}$). 
    \item For \textit{DW-2}, the HHCC-vs.-$U_0$ dependence was recorded as well, but between $-10$ and $+10$~V and -- due to the lower current level -- only up to the fourth harmonic order ($\omega_1/2 \pi = \SI{38.5}{Hz}$, $U_1 = \SI{0.4}{V}$).
    \item Furthermore, for \textit{DW-2} the HHCCs as a function of the excitation amplitude $U_1$ between \SI{10}{mV} and \SI{3}{V} with a constant offset voltage of $U_0 = \SI{0.7}{V}$ and a frequency of $\omega_1/2 \pi=23$~Hz  up to the fourth harmonic order has been acquired.
\end{itemize}

\subsection{Mathematical background of the HHCC analysis} 
\label{sec:methods:math}

The following considerations aim at introducing the relation between a static I-V curve and the corresponding HHCCs, demonstrating the additional value of the latter ones for electronic device characterization.
While there is an AC excitation voltage $U(t)$ of \cref{equ:exc_voltage} applied to the sample, we assume the latter to exhibit a purely static current response $I(t)$, excluding for instance capacitive and inductive contributions that can be suppressed in most cases by lowering the frequency. Pictorially, the systems moves, driven by the alternating excitation voltage, periodically forward and backward on the I-V curve (as sketched in \cref{fig:ac_principle:IVcurve}) that still determines the instantaneous value of the electric current.

Due to the fixed periodicity of the electric current response $I(t)$, given by the angular frequency of the excitation voltage $\omega_1$, it can be decomposed into higher harmonics of the excitation frequency within a Fourier transformation as:
\begin{equation}
    I(t) = \frac{-i}{2 \pi} \left[ I_0 + \sum_{m = 1}^{\infty} \left( I_m \exp(i \omega_m t)
        - \overline{I_m} \exp(- i \omega_m t) \right) \right],
\end{equation}
with the (complex) Fourier coefficients, introduced earlier as higher harmonic current contributions (HHCC), given by:
\begin{equation}
    I_m = i \int_0^{2 \pi/\omega_1}  I_\text{DC}(U_0 + U_1 \sin(i \omega_1 t)) \, \exp(- i m \omega_1 t) \text{d}t \quad .
    \label{equ:fourier_integral}
\end{equation}
The applied unusual sign convention is chosen intentionally to simplify the later comparison with experimental data. Several aspects motivate to investigate these coefficients $I_m$ in more detail:
\begin{enumerate}
    \item The definition of the coefficients [\cref{equ:fourier_integral}], i.e., multiplying the input signal with a harmonic reference signal and low-pass filtering by time integration, perfectly co-aligns with the working principle of a lock-in amplifier. This means that these coefficients are experimentally easily accessible by the latter devices and supported by the capability of certain lock-in amplifiers to generate the higher harmonic reference signals internally.
    \item The HHCCs are highly sensitive to the "fine structure" of the underlying  I-V curve. This is best illustrated by considering the local Taylor expansion of the I-V curve, representing the first and higher-order derivatives of the I-V curve around a certain bias voltage $U_0$. As shown in sec.~B of the Supplemental Material \cite{supplement}, under reasonable assumptions the coefficient $I_m$ is proportional to the $m^\text{th}$ derivative of the I-V curve, so it precisely characterizes the curve and may act as a key tool to identify the present transport mechanisms.
    \item \Cref{equ:fourier_integral} facilitates to \emph{predict} theoretical HHCCs based on given I-V curves (corresponding to different conduction models) and compare them to the respective HHCC measurements. This provides an independent and more precise way to identify transport mechanisms beyond DC I-V curve fitting, particularly for the case that different conduction models seem to fit the static I-V curve nearly equally well.
\end{enumerate}

In case of very simple conduction models, the integral in \cref{equ:fourier_integral} has an analytical solution that in turn allows for analytical predictions of the coefficients $I_m$. Apart from polynomial models like Ohmic transport and space charge limited conduction that can be solved using an integral table, another notable example is the Hopping transport, where the I-V characteristic is given by the Shockley equation (see also \cref{tab:conduction_mechanisms}) and the HHCCs are -- calculated in sec.~B of the Supplemental Material \cite{supplement} -- given by:
    
\begin{equation}
   I_m = I_\text{HT} \cdot
       \mathcal{I}_m \left( \frac{U_1}{U_\text{HT}} \right)
       \exp \left( \frac{U_0}{U_\text{HT}} \right) \exp \left(
        - i \frac{m - 1}{2} \pi \right).
   \label{equ:diode_fourier_coefficients}
\end{equation}

Here, \(I_\text{HT}\) is the diode's saturation current and \(U_\text{HT}\) the characteristic voltage, also expressed as \(U_\text{HT} = n k_\text{B} T / q\) with \(n\) being the ideality factor, \(k_\text{B}\) the Boltzmann constant, \(T\) the temperature, and \(q\) the elementary charge. \(\mathcal{I}_m(\cdot)\) denotes the modified Bessel function of first kind and \(m^\text{th}\) order. \Cref{equ:diode_fourier_coefficients} predicts the HHCC coefficients as a function of amplitude \(U_1\) and offset voltage \(U_0\) that can be compared to experimental results (cf., in particular, the reference data in sec.~D of the Supplemental Material \cite{supplement}, recorded on a commercial Schottky diode). The HHCC's real and imaginary parts of different orders $m$ form a characteristic pattern in the complex plane (Nyquist diagram) that is visualized in \cref{fig:ac_principle:complex_plane} and discussed in the following section.\\
To predict HHCCs for advanced and composite models as are the \textit{R2X2} models, a numerical solution of \cref{equ:fourier_integral} is required. This is achieved by combining a standard integration algorithm with the numerical evaluation of composite I-V curves described in  \cref{sec:iv_fits}. The procedure was applied to the specific cases of \textit{DW-1} and \textit{DW-2} and the results are presented in \cref{sec:results:HHCC_R2X2}.

\subsection{Method validation and consistency checks of the HHCC analysis}
\label{sec:results:consistency}

\begin{figure*}[htb!]
	\centering
	\includegraphics[width=\textwidth]{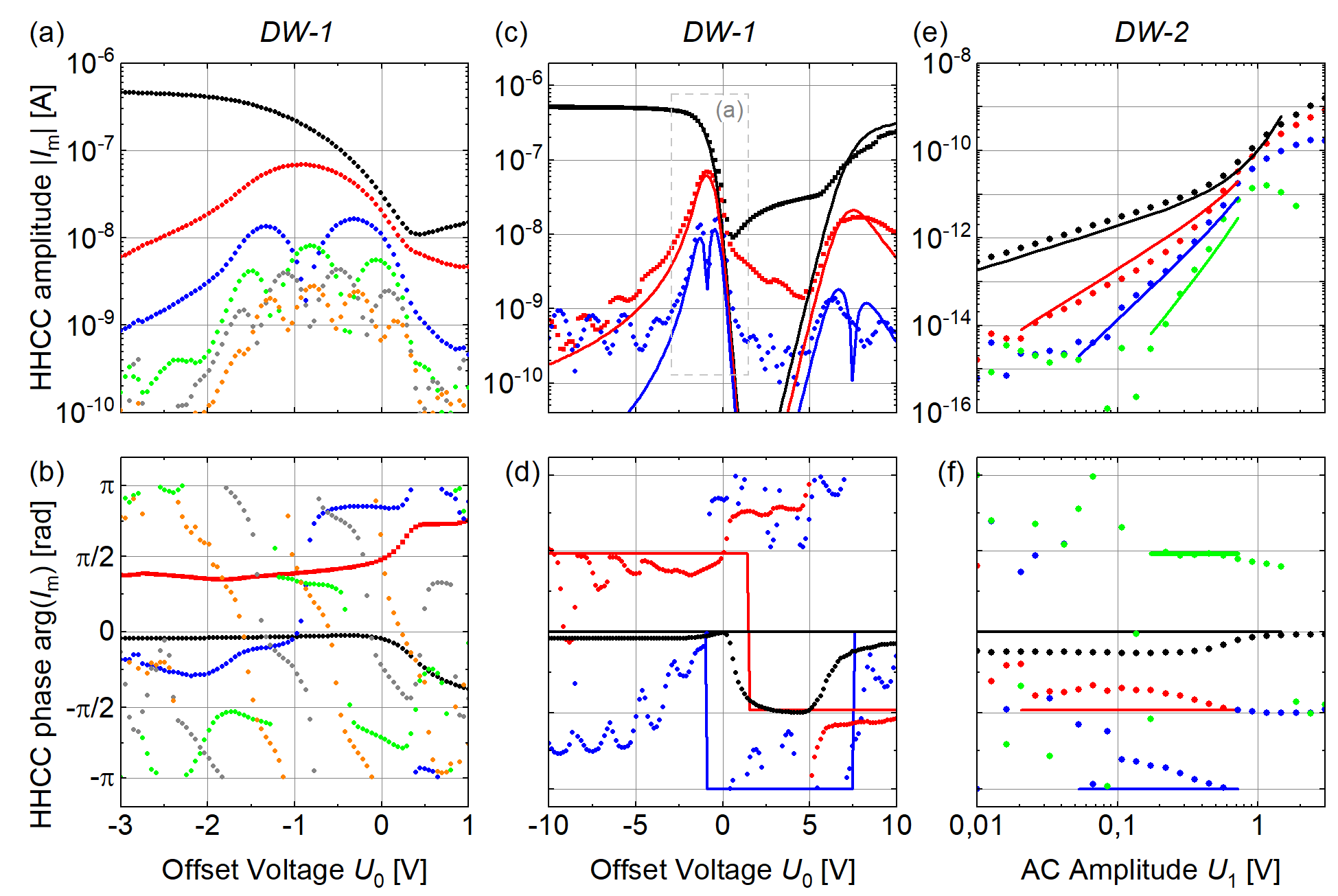}
	{\phantomsubfloat{\label{fig:ac_results:offset-detail:amp}}
     \phantomsubfloat{\label{fig:ac_results:offset-detail:phase}}
     \phantomsubfloat{\label{fig:ac_results:offset:amp}}
     \phantomsubfloat{\label{fig:ac_results:offset:phase}}	 \phantomsubfloat{\label{fig:ac_results:amplitude:amp}}
     \phantomsubfloat{\label{fig:ac_results:amplitude:phase}}}
     \vspace{4mm}
	\caption{AC conductance of \ce{MgO{:}LiNbO3} domain walls, represented as absolute value \(\lvert I_m \rvert\) (upper panels) and phase angle \(\arg(I_m)\) (lower panels) of higher harmonic current contributions~(HHCCs). The color coding for the harmonic orders $m$ is the same as in \cref{fig:ac_principle:complex_plane}. Experimental data is illustrated with dots, while theoretical predictions based on the I-V curve of the best-fitting \textit{R2D2} model are shown with solid lines.
    \textbf{(a), (b)} HHCC amplitude and phase as a function of the DC offset voltage on sample \textit{DW-1} (constant excitation parameters: $\omega_1/2 \pi = \SI{1.5}{kHz}$, $U_1 = \SI{0.71}{V}$) within a small voltage range. They fulfill several relations discussed in \cref{sec:results:consistency}, acting as consistency checks for the measurement working principle.
    \textbf{(c), (d)} Full-range DC offset voltage dependence, extending the view shown in panel (a) and (b) that is indicated in gray in the panel (c). A good agreement with the \textit{R2D2} model is observed for the amplitudes and phases of the first to third harmonic order, as discussed in mich more detail in the text.
    \textbf{(e), (f)} HHCCs of sample \textit{DW-2} under variable AC amplitude (constant parameters: $\omega_1/2 \pi = \SI{23}{\hertz}$, $U_0 = \SI{0.7}{V}$).}
	\label{fig:ac_results}
\end{figure*}

Before we proceed with the measurement of \ce{LiNbO3} domain walls' HHCCs and their comparison with the HHCCs \emph{predictions} derived from the several I-V characteristics' fitting models of \cref{tab:conduction_mechanisms}, we report on several stages of preliminary tests to validate the HHCC acquisition setup, since this type of measurement bears a number of pitfalls, such as harmonic distortions by the experimental setup or capacitive response contributions by the sample. Readers, who are primarily interested in the final derivation of the best-fitting transport process within the DW-electrode interfaces, may immediately jump to \cref{sec:results}.

First, the setup was tested on a \textit{commercial Schottky diode} (see sec.~D of the Supplemental Material \cite{supplement}) by recording the HHCCs as a function of the AC excitation amplitude \(U_1\) and the DC offset voltage \(U_0\). Both the modified-Bessel-function dependence of the HHCC amplitudes \(\abs{I_m}\) with respect to \(U_1\) and the exponential dependence with respect to \(U_0\) as given by \cref{equ:diode_fourier_coefficients} could be confirmed, proving both the high sensitivity for non-ohmic I-V features and the capability for quantitative comparisons within the mathematical framework described in \cref{sec:methods:math} and sec.~B of the Supplemental Material \cite{supplement}.

Second, the \textit{LNO bulk material}, which acts later as the in-situ reference (as sketched in \cref{fig:ac_principle:circuit}), has been analyzed separately with an adopted circuit without a reference path. As evaluated in sec.~A of the Supplemental Material \cite{supplement}, it was confirmed that the bulk material behaves like a simple parallel-plate capacitor, showing neither real-part (first harmonic order) conductance nor HHCCs at all.

Third, further elaborate consistency checks were performed using the two DW/electrode structures \textit{DW-1} and \textit{DW-2}, which are the actual focus of this work. \Cref{fig:ac_results} contains an instructive selection of HHCC amplitudes \(\abs{I_m}\) (a,c,e) and phases \(\arg(I_m)\) (b,d,f) measured as a function of \(U_0\) for sample \textit{DW-1} (a--d) and of \(U_1\) for \textit{DW-2} (e,f), which will be discussed with respect to the methodology in the following.

In particular, \cref{fig:ac_results:offset-detail:amp} and \subref{fig:ac_results:offset-detail:phase}, which are close-up views of \cref{fig:ac_results:offset:amp} and \cref{fig:ac_results:offset:phase}, show the HHCCs for \(m = 1\) to 6 captured for sample \textit{DW-1} within a selected range of DC offset voltages \(U_0\). The dataset reveals several non-trivial observations, which convincingly confirm the measurement principle for a complex structure as a \ce{LiNbO3} domain wall:

    (I): As mathematically shown in sec.~B of the Supplemental Material \cite{supplement}, in case of sufficiently-weak nonlinearity (as defined there), the current contribution of each harmonic order, \(I_m\), is proportional to the \(m^\text{th}\) derivative of the I-V characteristic \(I_\text{DC}(U)\). $I_m$ of a given order $m$ as a function of $U_0$ is therefore proportional to the derivative of the previous order's contribution $I_{m-1}$:
    
    \begin{equation}
        \frac{\text{d}}{\text{d} U_0} I_{m - 1} \propto \frac{\text{d}}{\text{d} U_0} \frac{\text{d}^{m - 1} I_\text{DC}}{\text{d} {U_\text{DC}}^{m - 1}}
            \underset{U_0 = U_\text{DC}}{=} \frac{\text{d}^{m} I_\text{DC}}{\text{d} {U_\text{DC}}^{m}} \propto I_{m} \quad ,
    \end{equation}
    
    This phenomenon is clearly evident at the local maxima of the amplitudes $\abs{I_m}$ depicted in \cref{fig:ac_results:offset-detail:amp}, which occur at the inflection points (maximum slope) of the preceding order's curve.
    
   (II): As derived more generally in sec.~B of the Supplemental Material \cite{supplement} and also applied to the special case of the hopping transport (HT) in \cref{equ:diode_fourier_coefficients}, neighboring HHCC orders ($I_m \rightarrow I_{m + 1}$) exhibit a phase difference of \SI{90}{\degree}, and, more specifically for the diode case, the rotation in the complex plane is strictly counter-clockwise (see also the theoretical Nyquist diagram in \cref{fig:ac_principle:complex_plane}). The first relation is observed for the phase (\cref{fig:ac_results:offset-detail:phase}) over a wide range, while a strict rotation is found, e.g., at \(U_0 = 0\) in clockwise direction due to the dominant backward diode. ($1^\text{st}$ order (black) $\rightarrow$ \SI{0}{\degree}, $2^\text{nd}$ order (red) $\rightarrow$ \SI{+90}{\degree} \textit{etc.}).
   
    (III): Moving beyond the "single-diode/hopping-transport" case, the HHCCs can vanish or, in other words, cross the origin in the complex plane, for certain combinations of $(m, U_0, U_1)$, as elaborated in sec.~B of the Supplemental Material \cite{supplement} [\cref{equ:app_math:hhcc_general}]. This phenomenon is indeed observed as sharp triangular-shaped minima of the amplitude (\cref{fig:ac_results:offset-detail:amp}), which coincide with a \SI{180}{\degree} shift of the corresponding phase. Due to observation (I) these zero-crossings reoccur in higher orders.

Furthermore, the full-range \(U_0\)-dependence of \(I_m\) of \textit{DW-1} for the first three harmonic orders \(m = 1\) to 3 together with the theoretical predictions based on the \textit{R2D2}-model fit curves of the DC I-V characteristics from \cref{fig:lno_iu_curves} (solid lines), which are depicted in \cref{fig:ac_results:offset:amp} and \subref{fig:ac_results:offset:phase}, reveal a satisfying agreement between experimental and numerically calculated data. Generally, in any given region (offset-voltage range) of the I-V characteristics, the curve is dominated by one of the circuit elements of the \textit{R2D2} equivalent circuit and consequently the HHCCs follow approximately the behavior of this single circuit element as well. For example, at positive  voltages larger than \SI{+8}{V}, the I-V curve is dominated by the forward resistor $R_\text{f}$, while the backward resistor $R_\text{b}$ dominates at negative voltages below \SI{-3}{V}, creating a strong first harmonic order current signal and minor high-order HHCCs. The backward and forward diode dominate around \(U_0 \approx \SI{-1}{\volt}\) and \(U_0 \approx \SI{+7}{\volt}\), respectively, and generate the local maxima in the second and third order of the amplitude. Since all HHCCs are expected to drop by several orders of magnitude between \(U_0 = \SI{0.5}{\volt}\) and \SI{6}{V}, differences in capacity of the reference and signal electrodes become the dominant contribution leading to deviations from the model calculation \cite{zahn_nonlinear_2022}.

Finally, a similarly satisfying agreement of the experimental data and the respective model calculation including \textit{only one branch} of the \textit{R2D2} model is observed for sample \textit{DW-2}. Due to the dominant positive branch of the I-V curve (cf. again \cref{fig:lno_dc_dw2}) and the low offset voltage of $U_0 = \SI{0.7}{V}$ -- chosen on purpose to be \textit{within} the diode-dominated regime -- the AC amplitude dependence shown in \cref{fig:ac_results:amplitude:amp} is close to the modified Bessel function-like shape as predicted by \cref{equ:diode_fourier_coefficients} for a single diode. Clear deviations occur below the detection limit of \SI{e-14}{\ampere}. An obvious change of the slope is observed in the $\abs{I_m}$-vs.-$U_0$ curve indicating the characteristic voltage $U_\text{HT}$. The respective phases agree as well with the predicted counter-clockwise rotation according to \cref{equ:diode_fourier_coefficients}. 

In summary, the different types of test and reference measurements confirm the suitability of the HHCC acquisition setup for analyzing the specific \ce{LiNbO3} domain wall structures, which are in focus here. They also validate -- without loss of generality -- the previously proposed \textit{R2D2} model to describe non-ohmic domain-wall conductance in \ce{LiNbO3}. However, going beyond this result, the HHCC analysis will be used to compare potential alternative equivalent circuit models (ECMs), named \emph{R2X2} models before, with the \textit{R2D2} model in the following.

\section{Results and Discussion}
\label{sec:results}

We proceed with (i) in-depth fitting attempts of the static I-V characteristics (cf.~\cref{sec:iv_fits}) and (ii) the discussion of HHCC measurements (cf.~\cref{sec:hhcc_derivation_main}) in comparison to the HHCCs, which were predicted mathematically based on the fit parameters of the static I-V curves (cf.~\cref{sec:methods:math}) -- all with the final goal to find the best-suited "\textit{R2X2}" model for describing the transport within the Cr-electrode/lithium-niobate-DW structures of this work. In particular, four different models, i.e., hopping transport~(HT), space-charge limited conduction~(SCLC), thermionic emission~(TE), and Fowler-Nordheim tunneling~(FNT) (cf.~\cref{tab:conduction_mechanisms}) were tested for the non-ohmic part ("\textit{X}-part") of the generalized "\textit{R2X2}" equivalent circuit.

\subsection{I-V curve fits using different \texorpdfstring{\textit{R2X2}}{R2X2} equivalent circuit models}
\label{sec:results:dc_R2X2}

\begin{figure*}[htb]
    \centering
    \includegraphics[width=\textwidth]{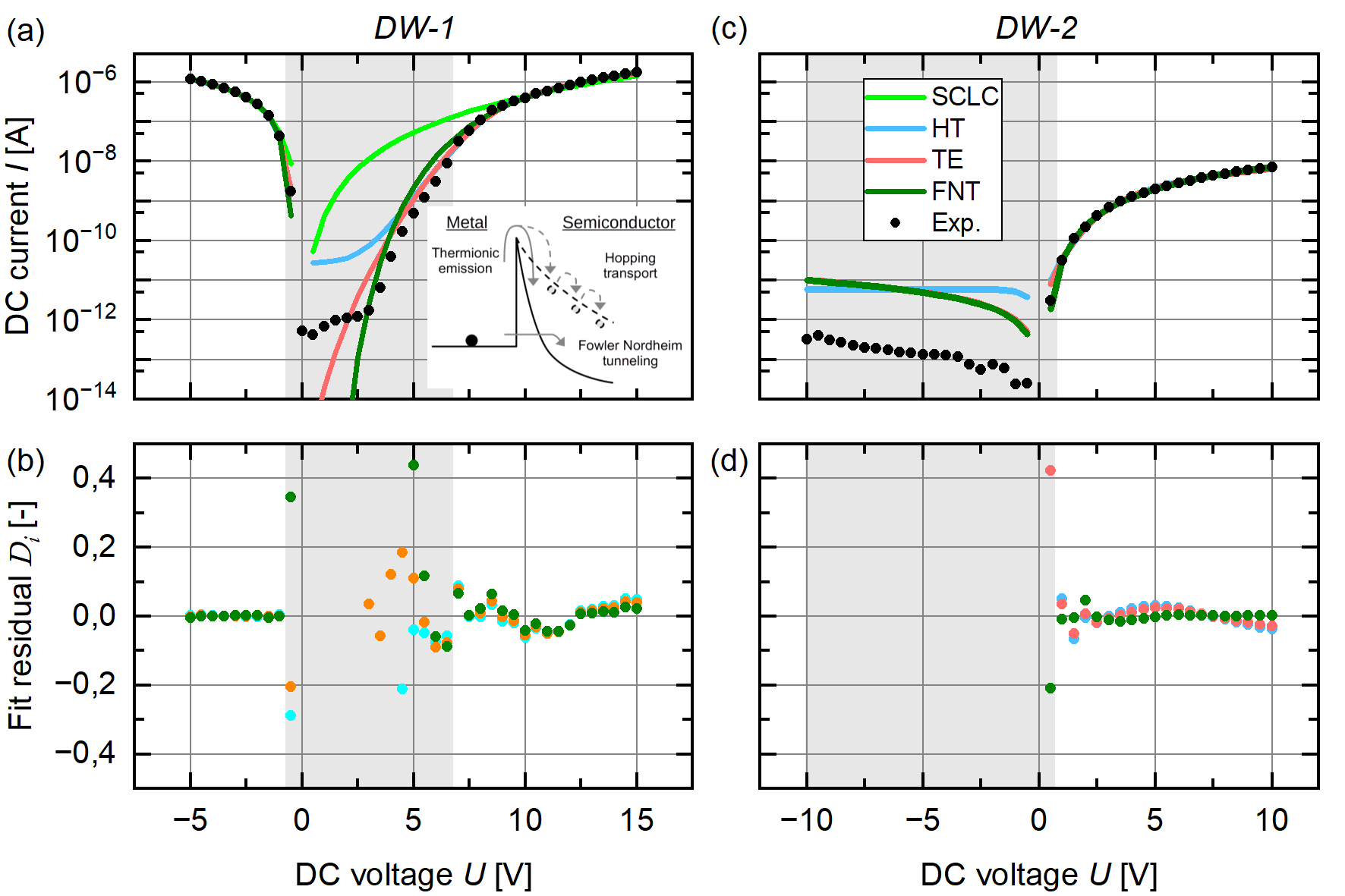}
    \vspace{2mm}
    {\phantomsubfloat{\label{fig:ac_models:iucurves:DW1-current}}
     \phantomsubfloat{\label{fig:ac_models:iucurves:DW1-deviation}}
     \phantomsubfloat{\label{fig:ac_models:iucurves:DW2-current}}
     \phantomsubfloat{\label{fig:ac_models:iucurves:DW2-deviation}}}
    \caption{Static I-V characteristics (recorded with $\text{d}V/\text{d}t = \SI{0.5}{V/s}$) of samples \textit{DW-1} \textbf{(a)} and \textit{DW-2} \textbf{(b)} modeled with different equivalent circuits of the \textit{R2X2} type. The space charge limited conduc\-tion~(SCLC) model (light green) is plotted next to the hopping transport (light blue), the thermionic emission (orange), and Fowler-Nordheim tunneling models (dark green). Since the latter three models are hard to distinguish visually, the residuals between experimental data and fit curves, are plotted in the the lower panels \textbf{(c)} and \textbf{(d)}. The grayish ranges were excluded from the fitting procedure. The inset within panel (a) sketches the three best-fitting \textit{X}-part processes within a simplified band scheme.}
    \label{fig:ac_models:iucurves}
\end{figure*}

\Cref{fig:ac_models:iucurves:DW1-current} depicts, as already shown in \cref{fig:lno_dc_dw1}, the measured static I-V curve of sample \emph{DW-1} as black dots, but now in logarithmic representation and in a shifted range: between \SI{-5}{V} and \SI{+15}{V}, to include a larger nonlinear region than available in the standard $\pm10$-V-interval for this specific sample. The four colored solid lines show the fitting results after applying the routine sketched in \cref{sec:iv_fits} using the \textit{R2X2} circuit model with the mechanisms of HT (light blue), SCLC (light green), TE (orange), and FNT (dark green) representing the \textit{X}-part.

When comparing the fit curves with the experimental data \emph{purely visually} in a first step, the model using the SCLC mechanism obviously performs decisively worse than the other cases, while the latter three candidates (HT, TE, FNT) lie close together. Having \cref{tab:conduction_mechanisms} in mind, one might ask whether the PFE and the TFE models were excluded. The reason for omitting the PFE case is that it gives very similar results as the TE model, since both have the same argument of the exponential function and there are currently no indications that a Poole-Frenkel effect (field-assisted trap-to-trap hopping) would be the dominant transport mechanism in lithium niobate. The TFE model, which, as one of the interface-limited mechanisms, is physically  a kind of a hybrid between the Fowler-Nordheim and the thermionic emission process, performs decisively worse (cf. the corresponding $R^2$ and $\mathcal{D}$ values in SI-table~S2) and is therefore also not taken into account in all the following calculations. However, with the other three possible cases, it is clear that more elaborate evaluations appear to be necessary to pinpoint the best-fitting model.

For that purpose, the data-point-specific residuals $\mathcal{D}_i$, as defined in \cref{equ:residuals}, are calculated (except for the grayish ranges, where the absolute measurement uncertainty inhibits a reliable parameter optimization) and plotted in \cref{fig:ac_models:iucurves:DW1-deviation}, while the sum $\mathcal{D}$ of all these pointwise residuals is given in \cref{tab:main:residuals} (for the respective $R^2$ values of the fitting routine, see table~S2 of the Supplemental Material \cite{supplement}). As a result, formally the FNT model performs best, showing the lowest $\mathcal{D}$ and highest $R^2$ values. However, the differences to the HT and TE models are rather small, which motivates to proceed with the HHCC analysis for further clarification and verification, as will be done in \cref{sec:results:HHCC_R2X2}.

For comparison, the I-V data of sample \emph{DW-2} has been processed in a similar way as depicted in \cref{fig:ac_models:iucurves:DW2-current,fig:ac_models:iucurves:DW2-deviation} with the fit parameters listed in table~S3 of the Supplemental Material \cite{supplement}. For that case, the negative branch shows a very low current level -- that is why we used the simplified equivalent circuit of only \emph{one} series connection of a resistor and a diode (or other rectifying element) as it had been sketched in the inset of \cref{fig:lno_dc_dw2}, in other words the fitting was based on an \emph{"RX"} model only. While, considering the positive branch of the I-V characteristics, the RX-model fit curves with the HT-, TE-, and FNT-mechanisms representing the \textit{X}-part of the equivalent circuit reproduce the experimental data all similarly well at a first glance, with a slightly worse fitting of the TE model for low voltages (\cref{fig:ac_models:iucurves:DW2-current}), the residuals' plots (\cref{fig:ac_models:iucurves:DW2-deviation}) more clearly point towards the FNT mechanism being the most suitable one, which means the same conclusion as for \textit{DW-1}.

\subsection{Comparison of measured HHCCs with their predictions from different \texorpdfstring{\textit{R2X2}}{R2X2} models}
\label{sec:results:HHCC_R2X2}

\begin{figure*}[htb]
    \centering
    {\phantomsubfloat{\label{fig:ac_models:hhcc_comparison:1st-amp}}
     \phantomsubfloat{\label{fig:ac_models:hhcc_comparison:1st-dev}}
     \phantomsubfloat{\label{fig:ac_models:hhcc_comparison:2nd-amp}}
     \phantomsubfloat{\label{fig:ac_models:hhcc_comparison:2nd-dev}}
     \phantomsubfloat{\label{fig:ac_models:hhcc_comparison:3rd-amp}}
     \phantomsubfloat{\label{fig:ac_models:hhcc_comparison:3rd-dev}}}
    \includegraphics[width=\textwidth]{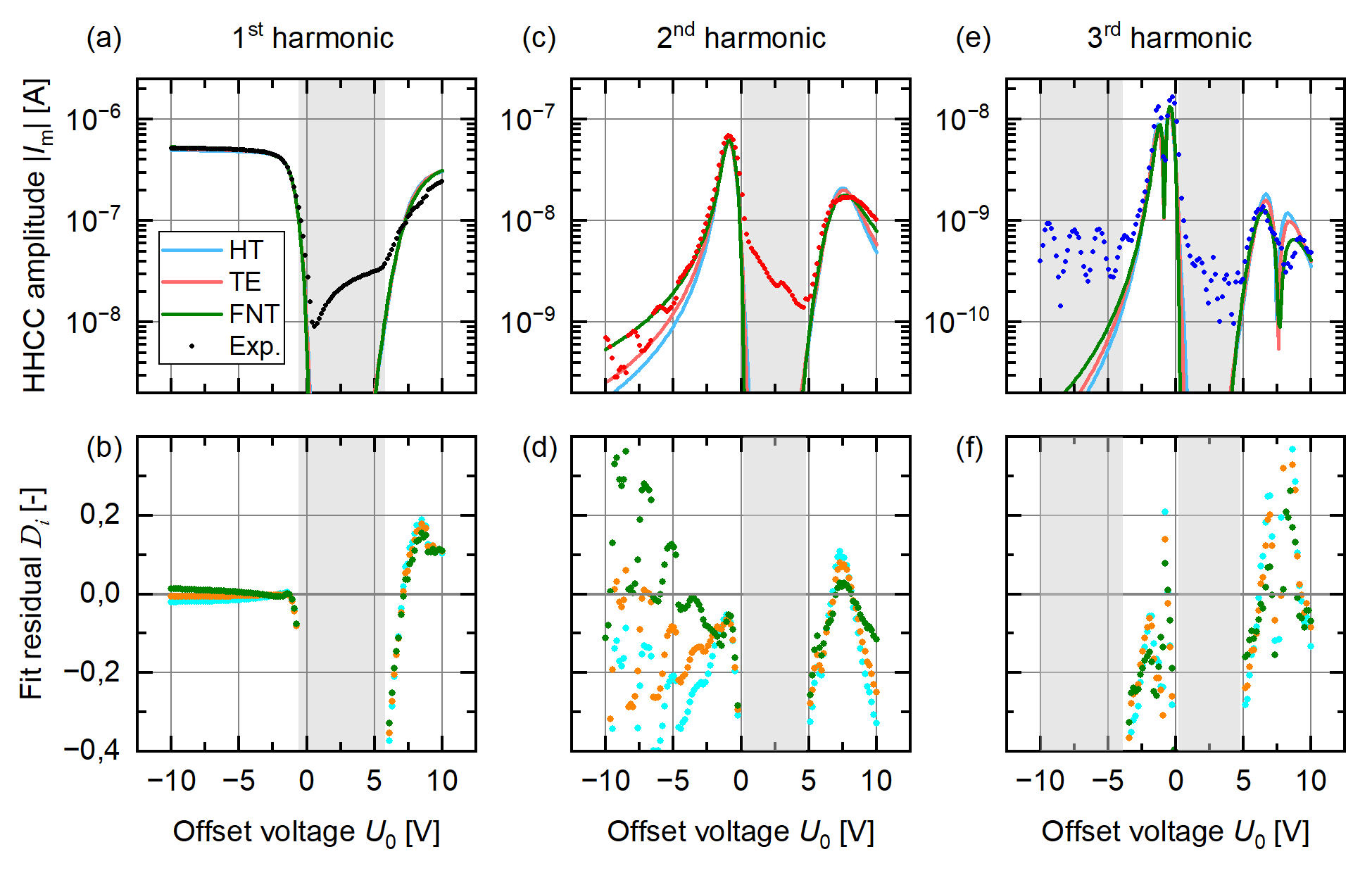}
    \caption{Comparison of measured and (on the basis of the static-I-V curve fit parameters) predicted/calculated HHCCs for sample \textit{DW-1}. The predictions are shown for the three best-performing equivalent-circuit models of the \textit{R2X2} type, using the hopping-transport/classical-diode description (light blue solid line), the thermionic emission model (orange solid line) and the Fowler-Nordheim tunneling (dark green solid line) for the \textit{X}-part. Separately shown are the measured absolute values of the \textbf{(a)} first (black dotted), \textbf{(c)} second (red dotted) and \textbf{(e)} third (blue dotted) harmonic order current contributions. Note that the experimental data and acquisition parameters are the same as in \cref{fig:ac_results:offset:amp}, that the complete data sets including also the HHCC phases and the weaker-performing SCLC and TFE models can be found in SI-fig.~S4, and that the grayish ranges were excluded from fitting. The lower panels \textbf{(b,d,f)} contain the corresponding residuals $\mathcal{D}_i$ for the three best-performing models showing that with rising harmonic order, it becomes clearer and clearer that the FNT model shows the lowest residuals.}
    \label{fig:ac_models:hhcc_comparison}
\end{figure*}

In order to verify and consolidate the conclusion drawn from the static-I-V characteristics' fits, we proceed with the evaluation of the higher-harmonic current contributions, measured up to the $3^\text{rd}$ (\textit{DW-1}) and $4^\text{th}$ (\textit{DW-2}) harmonic, as a function of the offset voltage as described in \cref{sec:hhcc_derivation_main}. In particular, we compare the "predicted" HHCC-vs.-$U_0$ curves, i.e., the functions, which were numerically calculated based on the different \textit{R2X2} fit parameters as obtained from the static I-V-curve fits (cf.~\cref{sec:results:dc_R2X2}), with the measured HHCCs, plot the residuals $\mathcal{D}_i$ as a function of $U_0$ and also calculate the residuals' sum $\mathcal{D}$ for those three \textit{R2X2} models ($X$=HT, TE, FNT), which were best-fitting but close together within the static-I-V fitting approach.

The corresponding results for sample \emph{DW-1} are shown for the first, second, and third harmonic in \cref{fig:ac_models:hhcc_comparison:1st-amp,fig:ac_models:hhcc_comparison:1st-dev}, (\subref{fig:ac_models:hhcc_comparison:2nd-amp}, \subref{fig:ac_models:hhcc_comparison:2nd-dev}), and (\subref{fig:ac_models:hhcc_comparison:3rd-amp}, \subref{fig:ac_models:hhcc_comparison:3rd-dev}), respectively, with the upper row panels showing measured HHCCs (dots) together with the 'predicted' functions using HT (light blue line), TE (orange line), and FNT (dark green line) representing the \textit{X}-part of the generalized \textit{R2X2} equivalent circuit model. Note that the measured HHCC amplitudes of \cref{fig:ac_models:hhcc_comparison} were intentionally already shown in \cref{fig:ac_results:offset:amp} in conjunction with the preceding discussion of the method's consistency. For the case of the first harmonic \cref{fig:ac_models:hhcc_comparison:1st-amp}, the HHCC predictions reproduce the negative branch below zero offset voltage very well and (apart from the gray range) the positive branch still quite satisfactory, but the three tested models lie extremely close together. Considering the corresponding fit deviations $\mathcal{D}_i$ in the panel below (\cref{fig:ac_models:hhcc_comparison:1st-dev}), there is no coherent conclusion to be drawn, the HT model performs obviously slightly worse than the other two over the full $U_0$ range, but for negative $U_0$ values, the TE model shows the lowest residuals, while for the positive branch the FNT mechanism seems to be slightly superior. Numerically the FNT mechanism exhibits the lowest residual sum $\mathcal{D}$, see \cref{tab:main:residuals}. Looking at the experimental data and HHCC predictions of the second- and third-harmonic cases (\cref{fig:ac_models:hhcc_comparison:2nd-amp,fig:ac_models:hhcc_comparison:3rd-amp}) the FNT mechanism beats the other two candidates much clearer both from the visual impression of the predicted-vs.-measured-curves comparison as well as from the $\mathcal{D}_i$ plots (\cref{fig:ac_models:hhcc_comparison:2nd-dev,fig:ac_models:hhcc_comparison:3rd-dev}).

For sample \emph{DW-2} -- though showing a decisively different I-V characteristics as discussed before, which is represented rather by a \emph{RX} than a \emph{R2X2} circuit -- the comparison of the measured $U_0$ dependence of HHCC amplitudes of the first and second harmonic order (higher orders give current contributions around and below the detection limit and cannot be considered further, see SI-Fig.~S5a) with the respective predicted curves points also towards the Fowler-Nordheim tunneling to be the best-fitting mechanism with the \emph{RX} equivalent circuit, which becomes especially clear from the evaluation of the second-harmonic current's data (see fig.~S5 of the Supplemental Material \cite{supplement} and \cref{tab:main:residuals}). 

\begin{table}
	{\renewcommand{\arraystretch}{1.3}
    \begin{tabular}{lccc} \hline \hline
        \thead{Sample \& Fit Case} & \thead{HT} & \thead{FNT} & \thead{TE} \\ \hline
        \textit{DW-1} DC & 0.0150 & 0.0127 & 0.0142 \\
        \textit{DW-2} DC & 0.0074 & 0.0015 & 0.0042 \\
        \textit{DW-1} AC $1^\text{st}$ order & 0.638 & 0.470 & 0.571 \\
        \textit{DW-1} AC $2^\text{nd}$ order & 5.658 & 1.894 & 2.674 \\
        \textit{DW-1} AC $3^\text{rd}$ order & 4.202 & 3.050 & 3.295 \\
        \textit{DW-2} AC $1^\text{st}$ order & 0.140 & 0.076 & 0.108 \\
        \textit{DW-2} AC $2^\text{nd}$ order & 6.765 & 0.608 & 4.138 \\ \hline \hline
    \end{tabular}}
    \caption{Summary of the curve fit residuals for the samples \textit{DW-1} and \textit{DW-2} and the various  DC and AC curve fit attempts. Only the three best-suited \textit{X}-parts of the R2X2 model, i.e., hopping transport (HT), thermionic emission (TE), and Fowler-Nordheim tunneling (FNT) are taken into account here.}
    \label{tab:main:residuals}
\end{table}

We close this section with a discussion on the physical meaning of Fowler-Nordheim tunneling being identified as the best-fitting model within the \textit{R2X2} equivalent-circuit concept for the description of the metal-electrode/LNO-DW system. For that purpose we recapitulate first that the \textit{X}-part contains the transport behavior across the metal/LNO interface, where an energy barrier (Schottky barrier) builds up. In classical semiconductor physics a number of different transport mechanisms in the vicinity of such a barrier have been described, cf. again \cref{tab:conduction_mechanisms} and classified as bulk- vs. interface-limited. To avoid confusion: "bulk" as used here is restricted to the range of the barrier in a band scheme (energy bands as a function of position) and is neither to be confounded with the bulk of the crystal nor the domain-wall regions deeper in the crystal with flat bands. In other words, a bulk-limited transport process across a heterojunction (here: Cr-electrode/LNO-DW) means that there are free states within the barrier and the carrier transport is realized in \emph{several} steps like hopping from site to site, while an interface-limited process needs only \emph{one} step, such as a tunneling event. Keeping this in mind, our present results are in sharp contrast to our older assumption in conjunction with the \textit{R2D2} model, which corresponded to the classical diode behavior, meaning that the carriers move via hopping inside the barrier region. However, the refined fitting and measurement efforts of this study point towards the inherently different interface-limited process of Fowler-Nordheim tunneling, meaning that the carriers are subject to quantum mechanical tunneling through the barrier. This implies a much thinner barrier than expected so far (typically a few nm for FNT instead of several 100~nm for HT), which would, viewed from the practical side, allow for much smaller devices and higher integration density. To which extent this result, which was derived for only two individual (and rather different) LNO domain walls, can be generalized, remains subject of future research, as well as the question whether the type of interface transport can be even engineered via the present preparation protocols or other stimuli.

\section{Summary and outlook}
\label{sec:conclusion}

The present work was dedicated to an in-depth investigation of the electrical transport through conductive ferroelectric domain walls written into 5-mol\% MgO-doped lithium niobate single crystals and contacted with Cr electrodes using the example of two specimen with distinctly different DC I-V characteristics: one with a symmetric, one with a highly asymmetric I-V curve. Thereby, the former \textit{R2D2} equivalent circuit model, which was heuristically derived from the typical I-V curve shapes of these electrode/DW structures and assumes a parallel connection of two resistor/diode pairs, where the diodes represent the electrode/DW junctions on both crystal sides and the resistors the "bulk"-region of the conductive DWs, was generalized towards a "\textit{R2X2}" circuit model with "$X$" standing for other (either bulk- or interface-limited) transport models, replacing the diode-like circuit element. In particular and besides the formerly assumed hopping transport (HT), which was used in the earlier \textit{R2D2} approach, space-charge limited conduction~(SCLC), Poole-Frenkel emission (PFE), thermionic emission~(TE), thermionic field emission (TFE), and Fowler-Nordheim tunneling~(FNT) were taken into account. From fitting the static I-V characteristics, the SCLC, PFE, and TFE models were excluded due to an impossible differentiation from other models and low physical plausibility (PFE) or due to a decisively worse performance (SCLC, TFE) as compared to the other candidates. The HT, TE, and FNT models, however, performed similarly well, with only slightly lower residuals for the FNT case. Thus, a clear distinction of the performance of the three remaining models from DC IV-curve fitting appeared to be questionable. 

As a consequence, a higher-harmonic current contribution (HHCC) measurement setup, where the samples (i) are excited with an AC voltage within selected highly non-linear ranges of the I-V curve and (ii) the current response is analyzed by lock-in techniques, which mathematically corresponds to a Fourier analysis, up to the sixth harmonic order, was implemented, extensively tested, and finally used for the acquisition of the HHCCs as function of the offset voltage in the two LNO DWs under investigation. By comparison of these data sets with the mathematically predicted HHCC curves, as calculated from the respective Fourier integrals using the fit parameters from the static I-V curve fits, we could distinguish the performance of the models more clearly and found the Fowler-Nordheim tunneling to be indeed the best-fitting model of the \emph{R2X2}-model's \textit{X}-part for both samples, meaning that it is an interface-limited mechanism, which governs the transport across the energy barrier across the electrode/DW junction and not a bulk-limited one as assumed in earlier work. This suggests that the barrier is narrower than previously assumed and that electronic components based on such a barrier, which benefit from the spatial mobility of the ferroelectric domain walls as an additional degree of freedom, could be designed significantly smaller and thus realized in higher integration densities.

Whether this specific result holds for a larger variety of structures remains subject of future research, going along with the task of achieving a higher level of automation in the complex fitting procedure. However, this work shows the unique potential of HHCC measurement and analysis for a more precise characterization of nanoelectronic structures such as ferroelectric domain walls, where conventional I-V characteristics' analysis comes to its limits.

\section*{Acknowledgments}

We acknowledge technical support on the measurement scheme realization by Ralf Raupach. 
Financial support is acknowledged by the Deutsche Forschungsgemeinschaft (DFG, German
Research Foundation) through the CRC~1415 (ID: 417590517) and the FOR~5044 (ID: 426703838). L.M.E. acknowledges financial support by the Deutsche Forschungsgemeinschaft (DFG, German Research Foundation) through the W\"urzburg-Dresden Cluster of Excellence ctd.qmat -- Complexity, Topology, and Dynamics in Quantum Matter (EXC~2147, project-ID 390858490). M.Z. acknowledges funding from the German Academic Exchange Service via a Research Grant for Doctoral Students (ID: 91849816), the Studienstiftung des Deutschen Volkes via a Doctoral Grant and the Free State of Bavaria via a Marianne-Plehn scholarship. I.K.'s contribution to this project is co-funded by the European Union and co-financed from tax revenues on the basis of the budget adopted by the Saxon State Parliament.

\section*{Data Availability}

The data that support the findings of this article are not publicly available upon publication because it is not technically feasible and/or the cost of preparing, depositing, and hosting the data would be prohibitive within the terms of this research project. The data are available from the authors upon reasonable request.


\begin{thebibliography}{39}%
\makeatletter
\providecommand \@ifxundefined [1]{%
 \@ifx{#1\undefined}
}%
\providecommand \@ifnum [1]{%
 \ifnum #1\expandafter \@firstoftwo
 \else \expandafter \@secondoftwo
 \fi
}%
\providecommand \@ifx [1]{%
 \ifx #1\expandafter \@firstoftwo
 \else \expandafter \@secondoftwo
 \fi
}%
\providecommand \natexlab [1]{#1}%
\providecommand \enquote  [1]{``#1''}%
\providecommand \bibnamefont  [1]{#1}%
\providecommand \bibfnamefont [1]{#1}%
\providecommand \citenamefont [1]{#1}%
\providecommand \href@noop [0]{\@secondoftwo}%
\providecommand \href [0]{\begingroup \@sanitize@url \@href}%
\providecommand \@href[1]{\@@startlink{#1}\@@href}%
\providecommand \@@href[1]{\endgroup#1\@@endlink}%
\providecommand \@sanitize@url [0]{\catcode `\\12\catcode `\$12\catcode
  `\&12\catcode `\#12\catcode `\^12\catcode `\_12\catcode `\%12\relax}%
\providecommand \@@startlink[1]{}%
\providecommand \@@endlink[0]{}%
\providecommand \url  [0]{\begingroup\@sanitize@url \@url }%
\providecommand \@url [1]{\endgroup\@href {#1}{\urlprefix }}%
\providecommand \urlprefix  [0]{URL }%
\providecommand \Eprint [0]{\href }%
\providecommand \doibase [0]{https://doi.org/}%
\providecommand \selectlanguage [0]{\@gobble}%
\providecommand \bibinfo  [0]{\@secondoftwo}%
\providecommand \bibfield  [0]{\@secondoftwo}%
\providecommand \translation [1]{[#1]}%
\providecommand \BibitemOpen [0]{}%
\providecommand \bibitemStop [0]{}%
\providecommand \bibitemNoStop [0]{.\EOS\space}%
\providecommand \EOS [0]{\spacefactor3000\relax}%
\providecommand \BibitemShut  [1]{\csname bibitem#1\endcsname}%
\let\auto@bib@innerbib\@empty
\bibitem [{\citenamefont {Meier}\ and\ \citenamefont
  {Selbach}(2022)}]{meier_ferroelectric_2022}%
  \BibitemOpen
  \bibfield  {author} {\bibinfo {author} {\bibfnamefont {D.}~\bibnamefont
  {Meier}}\ and\ \bibinfo {author} {\bibfnamefont {S.~M.}\ \bibnamefont
  {Selbach}},\ }\bibfield  {title} {\bibinfo {title} {Ferroelectric domain
  walls for nanotechnology},\ }\href
  {https://doi.org/10.1038/s41578-021-00375-z} {\bibfield  {journal} {\bibinfo
  {journal} {Nature Reviews Materials}\ }\textbf {\bibinfo {volume} {7}},\
  \bibinfo {pages} {157} (\bibinfo {year} {2022})}\BibitemShut {NoStop}%
\bibitem [{\citenamefont {Mikolajick}\ \emph {et~al.}(2023)\citenamefont
  {Mikolajick}, \citenamefont {Park}, \citenamefont {{Begon-Lours}},\ and\
  \citenamefont {Slesazeck}}]{mikolajick_ferroelectric_2023}%
  \BibitemOpen
  \bibfield  {author} {\bibinfo {author} {\bibfnamefont {T.}~\bibnamefont
  {Mikolajick}}, \bibinfo {author} {\bibfnamefont {M.~H.}\ \bibnamefont
  {Park}}, \bibinfo {author} {\bibfnamefont {L.}~\bibnamefont
  {{Begon-Lours}}},\ and\ \bibinfo {author} {\bibfnamefont {S.}~\bibnamefont
  {Slesazeck}},\ }\bibfield  {title} {\bibinfo {title} {From {{Ferroelectric
  Material Optimization}} to {{Neuromorphic Devices}}},\ }\href
  {https://doi.org/10.1002/adma.202206042} {\bibfield  {journal} {\bibinfo
  {journal} {Advanced Materials}\ }\textbf {\bibinfo {volume} {35}},\ \bibinfo
  {pages} {2206042} (\bibinfo {year} {2023})}\BibitemShut {NoStop}%
\bibitem [{\citenamefont {{Everschor-Sitte}}\ \emph {et~al.}(2024)\citenamefont
  {{Everschor-Sitte}}, \citenamefont {Majumdar}, \citenamefont {Wolk},\ and\
  \citenamefont {Meier}}]{everschor-sitte_topological_2024}%
  \BibitemOpen
  \bibfield  {author} {\bibinfo {author} {\bibfnamefont {K.}~\bibnamefont
  {{Everschor-Sitte}}}, \bibinfo {author} {\bibfnamefont {A.}~\bibnamefont
  {Majumdar}}, \bibinfo {author} {\bibfnamefont {K.}~\bibnamefont {Wolk}},\
  and\ \bibinfo {author} {\bibfnamefont {D.}~\bibnamefont {Meier}},\ }\bibfield
   {title} {\bibinfo {title} {Topological magnetic and ferroelectric systems
  for reservoir computing},\ }\href
  {https://doi.org/10.1038/s42254-024-00729-w} {\bibfield  {journal} {\bibinfo
  {journal} {Nature Reviews Physics}\ }\textbf {\bibinfo {volume} {6}},\
  \bibinfo {pages} {455} (\bibinfo {year} {2024})}\BibitemShut {NoStop}%
\bibitem [{\citenamefont {Catalan}\ \emph {et~al.}(2012)\citenamefont
  {Catalan}, \citenamefont {Seidel}, \citenamefont {Ramesh},\ and\
  \citenamefont {Scott}}]{catalan_domain_2012}%
  \BibitemOpen
  \bibfield  {author} {\bibinfo {author} {\bibfnamefont {G.}~\bibnamefont
  {Catalan}}, \bibinfo {author} {\bibfnamefont {J.}~\bibnamefont {Seidel}},
  \bibinfo {author} {\bibfnamefont {R.}~\bibnamefont {Ramesh}},\ and\ \bibinfo
  {author} {\bibfnamefont {J.~F.}\ \bibnamefont {Scott}},\ }\bibfield  {title}
  {\bibinfo {title} {Domain wall nanoelectronics},\ }\href
  {https://doi.org/10.1103/RevModPhys.84.119} {\bibfield  {journal} {\bibinfo
  {journal} {Reviews of Modern Physics}\ }\textbf {\bibinfo {volume} {84}},\
  \bibinfo {pages} {119} (\bibinfo {year} {2012})}\BibitemShut {NoStop}%
\bibitem [{\citenamefont {Meier}(2015)}]{meier_functional_2015}%
  \BibitemOpen
  \bibfield  {author} {\bibinfo {author} {\bibfnamefont {D.}~\bibnamefont
  {Meier}},\ }\bibfield  {title} {\bibinfo {title} {Functional domain walls in
  multiferroics},\ }\href {https://doi.org/10.1088/0953-8984/27/46/463003}
  {\bibfield  {journal} {\bibinfo  {journal} {Journal of Physics: Condensed
  Matter}\ }\textbf {\bibinfo {volume} {27}},\ \bibinfo {pages} {463003}
  (\bibinfo {year} {2015})}\BibitemShut {NoStop}%
\bibitem [{\citenamefont {Sluka}\ \emph {et~al.}(2016)\citenamefont {Sluka},
  \citenamefont {Bednyakov}, \citenamefont {Yudin}, \citenamefont {Crassous},\
  and\ \citenamefont {Tagantsev}}]{sluka_charged_2016}%
  \BibitemOpen
  \bibfield  {author} {\bibinfo {author} {\bibfnamefont {T.}~\bibnamefont
  {Sluka}}, \bibinfo {author} {\bibfnamefont {P.}~\bibnamefont {Bednyakov}},
  \bibinfo {author} {\bibfnamefont {P.}~\bibnamefont {Yudin}}, \bibinfo
  {author} {\bibfnamefont {A.}~\bibnamefont {Crassous}},\ and\ \bibinfo
  {author} {\bibfnamefont {A.}~\bibnamefont {Tagantsev}},\ }\bibinfo {title}
  {Charged domain walls in ferroelectrics},\ in\ \href
  {https://doi.org/10.1007/978-3-319-25301-5_5} {\emph {\bibinfo {booktitle}
  {Topological Structures in Ferroic Materials: Domain Walls, Vortices and
  Skyrmions}}},\ \bibinfo {editor} {edited by\ \bibinfo {editor} {\bibfnamefont
  {J.}~\bibnamefont {Seidel}}}\ (\bibinfo  {publisher} {Springer International
  Publishing},\ \bibinfo {address} {Cham},\ \bibinfo {year} {2016})\ pp.\
  \bibinfo {pages} {103--138}\BibitemShut {NoStop}%
\bibitem [{\citenamefont {Bednyakov}\ \emph {et~al.}(2018)\citenamefont
  {Bednyakov}, \citenamefont {Sturman}, \citenamefont {Sluka}, \citenamefont
  {Tagantsev},\ and\ \citenamefont {Yudin}}]{bednyakov_physics_2018}%
  \BibitemOpen
  \bibfield  {author} {\bibinfo {author} {\bibfnamefont {P.~S.}\ \bibnamefont
  {Bednyakov}}, \bibinfo {author} {\bibfnamefont {B.~I.}\ \bibnamefont
  {Sturman}}, \bibinfo {author} {\bibfnamefont {T.}~\bibnamefont {Sluka}},
  \bibinfo {author} {\bibfnamefont {A.~K.}\ \bibnamefont {Tagantsev}},\ and\
  \bibinfo {author} {\bibfnamefont {P.~V.}\ \bibnamefont {Yudin}},\ }\bibfield
  {title} {\bibinfo {title} {Physics and applications of charged domain
  walls},\ }\href {https://doi.org/10.1038/s41524-018-0121-8} {\bibfield
  {journal} {\bibinfo  {journal} {npj Computational Materials}\ }\textbf
  {\bibinfo {volume} {4}},\ \bibinfo {pages} {1} (\bibinfo {year}
  {2018})}\BibitemShut {NoStop}%
\bibitem [{\citenamefont {Sharma}\ \emph {et~al.}(2019)\citenamefont {Sharma},
  \citenamefont {Schoenherr},\ and\ \citenamefont
  {Seidel}}]{sharma_functional_2019}%
  \BibitemOpen
  \bibfield  {author} {\bibinfo {author} {\bibfnamefont {P.}~\bibnamefont
  {Sharma}}, \bibinfo {author} {\bibfnamefont {P.}~\bibnamefont {Schoenherr}},\
  and\ \bibinfo {author} {\bibfnamefont {J.}~\bibnamefont {Seidel}},\
  }\bibfield  {title} {\bibinfo {title} {Functional ferroic domain walls for
  nanoelectronics},\ }\href {https://doi.org/10.3390/ma12182927} {\bibfield
  {journal} {\bibinfo  {journal} {Materials}\ }\textbf {\bibinfo {volume}
  {12}},\ \bibinfo {pages} {2927} (\bibinfo {year} {2019})}\BibitemShut
  {NoStop}%
\bibitem [{\citenamefont {Nataf}\ \emph {et~al.}(2020)\citenamefont {Nataf},
  \citenamefont {Guennou}, \citenamefont {Gregg}, \citenamefont {Meier},
  \citenamefont {Hlinka}, \citenamefont {Salje},\ and\ \citenamefont
  {Kreisel}}]{nataf_domain-wall_2020}%
  \BibitemOpen
  \bibfield  {author} {\bibinfo {author} {\bibfnamefont {G.}~\bibnamefont
  {Nataf}}, \bibinfo {author} {\bibfnamefont {M.}~\bibnamefont {Guennou}},
  \bibinfo {author} {\bibfnamefont {J.}~\bibnamefont {Gregg}}, \bibinfo
  {author} {\bibfnamefont {D.}~\bibnamefont {Meier}}, \bibinfo {author}
  {\bibfnamefont {J.}~\bibnamefont {Hlinka}}, \bibinfo {author} {\bibfnamefont
  {E.~K.~H.}\ \bibnamefont {Salje}},\ and\ \bibinfo {author} {\bibfnamefont
  {J.}~\bibnamefont {Kreisel}},\ }\bibfield  {title} {\bibinfo {title}
  {Domain-wall engineering and topological defects in ferroelectric and
  ferroelastic materials},\ }\href {https://doi.org/10.1038/s42254-020-0235-z}
  {\bibfield  {journal} {\bibinfo  {journal} {Nature Reviews Physics}\ }\textbf
  {\bibinfo {volume} {2}},\ \bibinfo {pages} {634} (\bibinfo {year}
  {2020})}\BibitemShut {NoStop}%
\bibitem [{\citenamefont {Meier}\ and\ \citenamefont
  {Selbach}(2021)}]{meier_ferroelectric_2021}%
  \BibitemOpen
  \bibfield  {author} {\bibinfo {author} {\bibfnamefont {D.}~\bibnamefont
  {Meier}}\ and\ \bibinfo {author} {\bibfnamefont {S.}~\bibnamefont
  {Selbach}},\ }\bibfield  {title} {\bibinfo {title} {Ferroelectric domain
  walls for nanotechnology},\ }\href
  {https://doi.org/10.1038/s41578-021-00375-z} {\bibfield  {journal} {\bibinfo
  {journal} {Nature Reviews Materials}\ }\textbf {\bibinfo {volume} {7}},\
  \bibinfo {pages} {157} (\bibinfo {year} {2021})}\BibitemShut {NoStop}%
\bibitem [{\citenamefont {Sharma}\ \emph {et~al.}(2022)\citenamefont {Sharma},
  \citenamefont {Moise}, \citenamefont {Colombo},\ and\ \citenamefont
  {Seidel}}]{sharma_roadmap_2022}%
  \BibitemOpen
  \bibfield  {author} {\bibinfo {author} {\bibfnamefont {P.}~\bibnamefont
  {Sharma}}, \bibinfo {author} {\bibfnamefont {T.~S.}\ \bibnamefont {Moise}},
  \bibinfo {author} {\bibfnamefont {L.}~\bibnamefont {Colombo}},\ and\ \bibinfo
  {author} {\bibfnamefont {J.}~\bibnamefont {Seidel}},\ }\bibfield  {title}
  {\bibinfo {title} {Roadmap for ferroelectric domain wall nanoelectronics},\
  }\href {https://doi.org/10.1002/adfm.202110263} {\bibfield  {journal}
  {\bibinfo  {journal} {Advanced Functional Materials}\ }\textbf {\bibinfo
  {volume} {32}},\ \bibinfo {pages} {2110263} (\bibinfo {year}
  {2022})}\BibitemShut {NoStop}%
\bibitem [{\citenamefont {Schaab}\ \emph {et~al.}(2018)\citenamefont {Schaab},
  \citenamefont {Skj{\ae}rv{\o}}, \citenamefont {Krohns}, \citenamefont {Dai},
  \citenamefont {Holtz}, \citenamefont {Cano}, \citenamefont {Lilienblum},
  \citenamefont {Yan}, \citenamefont {Bourret}, \citenamefont {Muller},
  \citenamefont {Fiebig}, \citenamefont {Selbach},\ and\ \citenamefont
  {Meier}}]{schaab_electrical_2018}%
  \BibitemOpen
  \bibfield  {author} {\bibinfo {author} {\bibfnamefont {J.}~\bibnamefont
  {Schaab}}, \bibinfo {author} {\bibfnamefont {S.~H.}\ \bibnamefont
  {Skj{\ae}rv{\o}}}, \bibinfo {author} {\bibfnamefont {S.}~\bibnamefont
  {Krohns}}, \bibinfo {author} {\bibfnamefont {X.}~\bibnamefont {Dai}},
  \bibinfo {author} {\bibfnamefont {M.~E.}\ \bibnamefont {Holtz}}, \bibinfo
  {author} {\bibfnamefont {A.}~\bibnamefont {Cano}}, \bibinfo {author}
  {\bibfnamefont {M.}~\bibnamefont {Lilienblum}}, \bibinfo {author}
  {\bibfnamefont {Z.}~\bibnamefont {Yan}}, \bibinfo {author} {\bibfnamefont
  {E.}~\bibnamefont {Bourret}}, \bibinfo {author} {\bibfnamefont {D.~A.}\
  \bibnamefont {Muller}}, \bibinfo {author} {\bibfnamefont {M.}~\bibnamefont
  {Fiebig}}, \bibinfo {author} {\bibfnamefont {S.~M.}\ \bibnamefont
  {Selbach}},\ and\ \bibinfo {author} {\bibfnamefont {D.}~\bibnamefont
  {Meier}},\ }\bibfield  {title} {\bibinfo {title} {Electrical half-wave
  rectification at ferroelectric domain walls},\ }\href
  {https://doi.org/10.1038/s41565-018-0253-5} {\bibfield  {journal} {\bibinfo
  {journal} {Nature Nanotechnology}\ }\textbf {\bibinfo {volume} {13}},\
  \bibinfo {pages} {1028} (\bibinfo {year} {2018})}\BibitemShut {NoStop}%
\bibitem [{\citenamefont {Niu}\ \emph {et~al.}(2023)\citenamefont {Niu},
  \citenamefont {Qiao}, \citenamefont {Lu}, \citenamefont {Fu}, \citenamefont
  {Liu}, \citenamefont {Bi}, \citenamefont {Mei}, \citenamefont {You},
  \citenamefont {Chou},\ and\ \citenamefont {Geng}}]{Niu_Diode_2023}%
  \BibitemOpen
  \bibfield  {author} {\bibinfo {author} {\bibfnamefont {L.}~\bibnamefont
  {Niu}}, \bibinfo {author} {\bibfnamefont {X.}~\bibnamefont {Qiao}}, \bibinfo
  {author} {\bibfnamefont {H.}~\bibnamefont {Lu}}, \bibinfo {author}
  {\bibfnamefont {W.}~\bibnamefont {Fu}}, \bibinfo {author} {\bibfnamefont
  {Y.}~\bibnamefont {Liu}}, \bibinfo {author} {\bibfnamefont {K.}~\bibnamefont
  {Bi}}, \bibinfo {author} {\bibfnamefont {L.}~\bibnamefont {Mei}}, \bibinfo
  {author} {\bibfnamefont {Y.}~\bibnamefont {You}}, \bibinfo {author}
  {\bibfnamefont {X.}~\bibnamefont {Chou}},\ and\ \bibinfo {author}
  {\bibfnamefont {W.}~\bibnamefont {Geng}},\ }\bibfield  {title} {\bibinfo
  {title} {{Diode-Like Behavior Based on Conductive Domain Wall in LiNbO
  Ferroelectric Single-Crystal Thin Film}},\ }\href
  {https://doi.org/10.1109/LED.2022.3224915} {\bibfield  {journal} {\bibinfo
  {journal} {IEEE Electron Device Letters}\ }\textbf {\bibinfo {volume} {44}},\
  \bibinfo {pages} {52} (\bibinfo {year} {2023})}\BibitemShut {NoStop}%
\bibitem [{\citenamefont {Jiang}\ \emph {et~al.}(2020)\citenamefont {Jiang},
  \citenamefont {Geng}, \citenamefont {Lv}, \citenamefont {wang Hong},
  \citenamefont {Jiang}, \citenamefont {Wang}, \citenamefont {Chai},
  \citenamefont {Lian}, \citenamefont {Zhang}, \citenamefont {Huang},
  \citenamefont {Zhang}, \citenamefont {Scott},\ and\ \citenamefont
  {Hwang}}]{Jiang_Ferroelectric_2020}%
  \BibitemOpen
  \bibfield  {author} {\bibinfo {author} {\bibfnamefont {A.~Q.}\ \bibnamefont
  {Jiang}}, \bibinfo {author} {\bibfnamefont {W.~P.}\ \bibnamefont {Geng}},
  \bibinfo {author} {\bibfnamefont {P.}~\bibnamefont {Lv}}, \bibinfo {author}
  {\bibfnamefont {J.}~\bibnamefont {wang Hong}}, \bibinfo {author}
  {\bibfnamefont {J.}~\bibnamefont {Jiang}}, \bibinfo {author} {\bibfnamefont
  {C.}~\bibnamefont {Wang}}, \bibinfo {author} {\bibfnamefont {X.~J.}\
  \bibnamefont {Chai}}, \bibinfo {author} {\bibfnamefont {J.~W.}\ \bibnamefont
  {Lian}}, \bibinfo {author} {\bibfnamefont {Y.}~\bibnamefont {Zhang}},
  \bibinfo {author} {\bibfnamefont {R.}~\bibnamefont {Huang}}, \bibinfo
  {author} {\bibfnamefont {D.~W.}\ \bibnamefont {Zhang}}, \bibinfo {author}
  {\bibfnamefont {J.~F.}\ \bibnamefont {Scott}},\ and\ \bibinfo {author}
  {\bibfnamefont {C.~S.}\ \bibnamefont {Hwang}},\ }\bibfield  {title} {\bibinfo
  {title} {{Ferroelectric domain wall memory with embedded selector realized in
  LiNbO$_3$ single crystals integrated on Si wafers}},\ }\href
  {https://doi.org/10.1038/s41563-020-0702-z} {\bibfield  {journal} {\bibinfo
  {journal} {Nature Materials}\ }\textbf {\bibinfo {volume} {19}},\ \bibinfo
  {pages} {1188} (\bibinfo {year} {2020})}\BibitemShut {NoStop}%
\bibitem [{\citenamefont {Hu}\ \emph {et~al.}(2024)\citenamefont {Hu},
  \citenamefont {Shen}, \citenamefont {Sun}, \citenamefont {Li},\ and\
  \citenamefont {Jiang}}]{Hu_Cryogenic_2024}%
  \BibitemOpen
  \bibfield  {author} {\bibinfo {author} {\bibfnamefont {D.}~\bibnamefont
  {Hu}}, \bibinfo {author} {\bibfnamefont {B.~W.}\ \bibnamefont {Shen}},
  \bibinfo {author} {\bibfnamefont {J.}~\bibnamefont {Sun}}, \bibinfo {author}
  {\bibfnamefont {Y.~M.}\ \bibnamefont {Li}},\ and\ \bibinfo {author}
  {\bibfnamefont {A.~Q.}\ \bibnamefont {Jiang}},\ }\bibfield  {title} {\bibinfo
  {title} {Cryogenic ferroelectric linbo3 domain wall memory},\ }\href
  {https://doi.org/10.1109/LED.2023.3346891} {\bibfield  {journal} {\bibinfo
  {journal} {IEEE Electron Device Letters}\ }\textbf {\bibinfo {volume} {45}},\
  \bibinfo {pages} {380} (\bibinfo {year} {2024})}\BibitemShut {NoStop}%
\bibitem [{\citenamefont {Yu}\ \emph {et~al.}(2026)\citenamefont {Yu},
  \citenamefont {Tang}, \citenamefont {Wang}, \citenamefont {Wang},
  \citenamefont {Li}, \citenamefont {Shen}, \citenamefont {Zhang},\ and\
  \citenamefont {Jiang}}]{Yu_Adjustable_2026}%
  \BibitemOpen
  \bibfield  {author} {\bibinfo {author} {\bibfnamefont {X.}~\bibnamefont
  {Yu}}, \bibinfo {author} {\bibfnamefont {H.}~\bibnamefont {Tang}}, \bibinfo
  {author} {\bibfnamefont {X.}~\bibnamefont {Wang}}, \bibinfo {author}
  {\bibfnamefont {Z.}~\bibnamefont {Wang}}, \bibinfo {author} {\bibfnamefont
  {Y.}~\bibnamefont {Li}}, \bibinfo {author} {\bibfnamefont {B.}~\bibnamefont
  {Shen}}, \bibinfo {author} {\bibfnamefont {W.~D.}\ \bibnamefont {Zhang}},\
  and\ \bibinfo {author} {\bibfnamefont {A.~Q.}\ \bibnamefont {Jiang}},\
  }\bibfield  {title} {\bibinfo {title} {{Adjustable embedded selectors in a
  ferroelectric LiNbO$_3$ domain-wall memory}},\ }\href
  {https://doi.org/10.1063/5.0315123} {\bibfield  {journal} {\bibinfo
  {journal} {Journal of Applied Physics}\ }\textbf {\bibinfo {volume} {139}},\
  \bibinfo {pages} {064103} (\bibinfo {year} {2026})}\BibitemShut {NoStop}%
\bibitem [{\citenamefont {Chai}\ \emph {et~al.}(2020)\citenamefont {Chai},
  \citenamefont {Jiang}, \citenamefont {Zhang}, \citenamefont {Hou},
  \citenamefont {Meng}, \citenamefont {Wang}, \citenamefont {Gu}, \citenamefont
  {Zhang},\ and\ \citenamefont {Jiang}}]{chai_nonvolatile_2020}%
  \BibitemOpen
  \bibfield  {author} {\bibinfo {author} {\bibfnamefont {X.}~\bibnamefont
  {Chai}}, \bibinfo {author} {\bibfnamefont {J.}~\bibnamefont {Jiang}},
  \bibinfo {author} {\bibfnamefont {Q.}~\bibnamefont {Zhang}}, \bibinfo
  {author} {\bibfnamefont {X.}~\bibnamefont {Hou}}, \bibinfo {author}
  {\bibfnamefont {F.}~\bibnamefont {Meng}}, \bibinfo {author} {\bibfnamefont
  {J.}~\bibnamefont {Wang}}, \bibinfo {author} {\bibfnamefont {L.}~\bibnamefont
  {Gu}}, \bibinfo {author} {\bibfnamefont {D.~W.}\ \bibnamefont {Zhang}},\ and\
  \bibinfo {author} {\bibfnamefont {A.~Q.}\ \bibnamefont {Jiang}},\ }\bibfield
  {title} {\bibinfo {title} {Nonvolatile ferroelectric field-effect
  transistors},\ }\href {https://doi.org/10.1038/s41467-020-16623-9} {\bibfield
   {journal} {\bibinfo  {journal} {Nature Communications}\ }\textbf {\bibinfo
  {volume} {11}},\ \bibinfo {pages} {2811} (\bibinfo {year}
  {2020})}\BibitemShut {NoStop}%
\bibitem [{\citenamefont {Seufert}\ \emph {et~al.}(2021)\citenamefont
  {Seufert}, \citenamefont {Hassanpour~Amiri}, \citenamefont {Gkoupidenis},\
  and\ \citenamefont {Asadi}}]{seufert_crossbar_2021}%
  \BibitemOpen
  \bibfield  {author} {\bibinfo {author} {\bibfnamefont {L.}~\bibnamefont
  {Seufert}}, \bibinfo {author} {\bibfnamefont {M.}~\bibnamefont
  {Hassanpour~Amiri}}, \bibinfo {author} {\bibfnamefont {P.}~\bibnamefont
  {Gkoupidenis}},\ and\ \bibinfo {author} {\bibfnamefont {K.}~\bibnamefont
  {Asadi}},\ }\bibfield  {title} {\bibinfo {title} {Crossbar {{Array}} of
  {{Artificial Synapses Based}} on {{Ferroelectric Diodes}}},\ }\href
  {https://doi.org/10.1002/aelm.202100558} {\bibfield  {journal} {\bibinfo
  {journal} {Advanced Electronic Materials}\ }\textbf {\bibinfo {volume} {7}},\
  \bibinfo {pages} {2100558} (\bibinfo {year} {2021})}\BibitemShut {NoStop}%
\bibitem [{\citenamefont {Seidel}\ \emph {et~al.}(2009)\citenamefont {Seidel},
  \citenamefont {Martin}, \citenamefont {He}, \citenamefont {Zhan},
  \citenamefont {Chu}, \citenamefont {Rother}, \citenamefont {Hawkridge},
  \citenamefont {Maksymovych}, \citenamefont {Yu}, \citenamefont {Gajek},
  \citenamefont {Balke}, \citenamefont {Kalinin}, \citenamefont {Gemming},
  \citenamefont {Wang}, \citenamefont {Catalan}, \citenamefont {Scott},
  \citenamefont {Spaldin}, \citenamefont {Orenstein},\ and\ \citenamefont
  {Ramesh}}]{seidel_conduction_2009}%
  \BibitemOpen
  \bibfield  {author} {\bibinfo {author} {\bibfnamefont {J.}~\bibnamefont
  {Seidel}}, \bibinfo {author} {\bibfnamefont {L.~W.}\ \bibnamefont {Martin}},
  \bibinfo {author} {\bibfnamefont {Q.}~\bibnamefont {He}}, \bibinfo {author}
  {\bibfnamefont {Q.}~\bibnamefont {Zhan}}, \bibinfo {author} {\bibfnamefont
  {Y.-H.}\ \bibnamefont {Chu}}, \bibinfo {author} {\bibfnamefont
  {A.}~\bibnamefont {Rother}}, \bibinfo {author} {\bibfnamefont {M.~E.}\
  \bibnamefont {Hawkridge}}, \bibinfo {author} {\bibfnamefont {P.}~\bibnamefont
  {Maksymovych}}, \bibinfo {author} {\bibfnamefont {P.}~\bibnamefont {Yu}},
  \bibinfo {author} {\bibfnamefont {M.}~\bibnamefont {Gajek}}, \bibinfo
  {author} {\bibfnamefont {N.}~\bibnamefont {Balke}}, \bibinfo {author}
  {\bibfnamefont {S.~V.}\ \bibnamefont {Kalinin}}, \bibinfo {author}
  {\bibfnamefont {S.}~\bibnamefont {Gemming}}, \bibinfo {author} {\bibfnamefont
  {F.}~\bibnamefont {Wang}}, \bibinfo {author} {\bibfnamefont {G.}~\bibnamefont
  {Catalan}}, \bibinfo {author} {\bibfnamefont {J.~F.}\ \bibnamefont {Scott}},
  \bibinfo {author} {\bibfnamefont {N.~A.}\ \bibnamefont {Spaldin}}, \bibinfo
  {author} {\bibfnamefont {J.}~\bibnamefont {Orenstein}},\ and\ \bibinfo
  {author} {\bibfnamefont {R.}~\bibnamefont {Ramesh}},\ }\bibfield  {title}
  {\bibinfo {title} {Conduction at domain walls in oxide multiferroics},\
  }\href {https://doi.org/10.1038/nmat2373} {\bibfield  {journal} {\bibinfo
  {journal} {Nature Materials}\ }\textbf {\bibinfo {volume} {8}},\ \bibinfo
  {pages} {229} (\bibinfo {year} {2009})}\BibitemShut {NoStop}%
\bibitem [{\citenamefont {Werner}\ \emph {et~al.}(2017)\citenamefont {Werner},
  \citenamefont {Herr}, \citenamefont {Buse}, \citenamefont {Sturman},
  \citenamefont {Soergel}, \citenamefont {Razzaghi},\ and\ \citenamefont
  {Breunig}}]{werner_large_2017}%
  \BibitemOpen
  \bibfield  {author} {\bibinfo {author} {\bibfnamefont {C.~S.}\ \bibnamefont
  {Werner}}, \bibinfo {author} {\bibfnamefont {S.~J.}\ \bibnamefont {Herr}},
  \bibinfo {author} {\bibfnamefont {K.}~\bibnamefont {Buse}}, \bibinfo {author}
  {\bibfnamefont {B.}~\bibnamefont {Sturman}}, \bibinfo {author} {\bibfnamefont
  {E.}~\bibnamefont {Soergel}}, \bibinfo {author} {\bibfnamefont
  {C.}~\bibnamefont {Razzaghi}},\ and\ \bibinfo {author} {\bibfnamefont
  {I.}~\bibnamefont {Breunig}},\ }\bibfield  {title} {\bibinfo {title} {Large
  and accessible conductivity of charged domain walls in lithium niobate},\
  }\href {https://doi.org/10.1038/s41598-017-09703-2} {\bibfield  {journal}
  {\bibinfo  {journal} {Scientific Reports}\ }\textbf {\bibinfo {volume} {7}},\
  \bibinfo {pages} {9862} (\bibinfo {year} {2017})}\BibitemShut {NoStop}%
\bibitem [{\citenamefont {Puntigam}\ \emph {et~al.}(2022)\citenamefont
  {Puntigam}, \citenamefont {Altthaler}, \citenamefont {Ghara}, \citenamefont
  {Prodan}, \citenamefont {Tsurkan}, \citenamefont {Krohns}, \citenamefont
  {K{\'e}zsm{\'a}rki},\ and\ \citenamefont {Evans}}]{puntigam_strain_2022}%
  \BibitemOpen
  \bibfield  {author} {\bibinfo {author} {\bibfnamefont {L.}~\bibnamefont
  {Puntigam}}, \bibinfo {author} {\bibfnamefont {M.}~\bibnamefont {Altthaler}},
  \bibinfo {author} {\bibfnamefont {S.}~\bibnamefont {Ghara}}, \bibinfo
  {author} {\bibfnamefont {L.}~\bibnamefont {Prodan}}, \bibinfo {author}
  {\bibfnamefont {V.}~\bibnamefont {Tsurkan}}, \bibinfo {author} {\bibfnamefont
  {S.}~\bibnamefont {Krohns}}, \bibinfo {author} {\bibfnamefont
  {I.}~\bibnamefont {K{\'e}zsm{\'a}rki}},\ and\ \bibinfo {author}
  {\bibfnamefont {D.~M.}\ \bibnamefont {Evans}},\ }\bibfield  {title} {\bibinfo
  {title} {Strain {{Driven Conducting Domain Walls}} in a {{Mott Insulator}}},\
  }\href {https://doi.org/10.1002/aelm.202200366} {\bibfield  {journal}
  {\bibinfo  {journal} {Advanced Electronic Materials}\ }\textbf {\bibinfo
  {volume} {2022}},\ \bibinfo {pages} {2200366} (\bibinfo {year}
  {2022})}\BibitemShut {NoStop}%
\bibitem [{\citenamefont {Holstad}\ \emph {et~al.}(2018)\citenamefont
  {Holstad}, \citenamefont {Evans}, \citenamefont {Ruff}, \citenamefont
  {Sm{\aa}br{\aa}ten}, \citenamefont {Schaab}, \citenamefont {Tzschaschel},
  \citenamefont {Yan}, \citenamefont {Bourret}, \citenamefont {Selbach},
  \citenamefont {Krohns},\ and\ \citenamefont
  {Meier}}]{holstad_electronic_2018}%
  \BibitemOpen
  \bibfield  {author} {\bibinfo {author} {\bibfnamefont {T.~S.}\ \bibnamefont
  {Holstad}}, \bibinfo {author} {\bibfnamefont {D.~M.}\ \bibnamefont {Evans}},
  \bibinfo {author} {\bibfnamefont {A.}~\bibnamefont {Ruff}}, \bibinfo {author}
  {\bibfnamefont {D.~R.}\ \bibnamefont {Sm{\aa}br{\aa}ten}}, \bibinfo {author}
  {\bibfnamefont {J.}~\bibnamefont {Schaab}}, \bibinfo {author} {\bibfnamefont
  {C.}~\bibnamefont {Tzschaschel}}, \bibinfo {author} {\bibfnamefont
  {Z.}~\bibnamefont {Yan}}, \bibinfo {author} {\bibfnamefont {E.}~\bibnamefont
  {Bourret}}, \bibinfo {author} {\bibfnamefont {S.~M.}\ \bibnamefont
  {Selbach}}, \bibinfo {author} {\bibfnamefont {S.}~\bibnamefont {Krohns}},\
  and\ \bibinfo {author} {\bibfnamefont {D.}~\bibnamefont {Meier}},\ }\bibfield
   {title} {\bibinfo {title} {Electronic bulk and domain wall properties in
  \(b\)-site doped hexagonal \ce{ErMnO3}},\ }\href
  {https://doi.org/10.1103/PhysRevB.97.085143} {\bibfield  {journal} {\bibinfo
  {journal} {Physical Review B}\ }\textbf {\bibinfo {volume} {97}},\ \bibinfo
  {pages} {085143} (\bibinfo {year} {2018})}\BibitemShut {NoStop}%
\bibitem [{\citenamefont {Qi}\ \emph {et~al.}(2015)\citenamefont {Qi},
  \citenamefont {Martirez}, \citenamefont {Saidi}, \citenamefont {Urban},
  \citenamefont {Yun}, \citenamefont {Spanier},\ and\ \citenamefont
  {Rappe}}]{qi_modified_2015}%
  \BibitemOpen
  \bibfield  {author} {\bibinfo {author} {\bibfnamefont {Y.}~\bibnamefont
  {Qi}}, \bibinfo {author} {\bibfnamefont {J.~M.~P.}\ \bibnamefont {Martirez}},
  \bibinfo {author} {\bibfnamefont {W.~A.}\ \bibnamefont {Saidi}}, \bibinfo
  {author} {\bibfnamefont {J.~J.}\ \bibnamefont {Urban}}, \bibinfo {author}
  {\bibfnamefont {W.~S.}\ \bibnamefont {Yun}}, \bibinfo {author} {\bibfnamefont
  {J.~E.}\ \bibnamefont {Spanier}},\ and\ \bibinfo {author} {\bibfnamefont
  {A.~M.}\ \bibnamefont {Rappe}},\ }\bibfield  {title} {\bibinfo {title}
  {Modified schottky emission to explain thickness dependence and slow
  depolarization in \ce{BaTiO3} nanowires},\ }\href
  {https://doi.org/10.1103/PhysRevB.91.245431} {\bibfield  {journal} {\bibinfo
  {journal} {Physical Review B}\ }\textbf {\bibinfo {volume} {91}},\ \bibinfo
  {pages} {245431} (\bibinfo {year} {2015})}\BibitemShut {NoStop}%
\bibitem [{\citenamefont {Nagasawa}\ and\ \citenamefont
  {Nozawa}(1999)}]{nagasawa_imprint_1999}%
  \BibitemOpen
  \bibfield  {author} {\bibinfo {author} {\bibfnamefont {D.}~\bibnamefont
  {Nagasawa}}\ and\ \bibinfo {author} {\bibfnamefont {H.}~\bibnamefont
  {Nozawa}},\ }\bibfield  {title} {\bibinfo {title} {Imprint {{Model Based}} on
  {{Thermionic Electron Emission Under Local Fields}} in {{Ferroelectric Thin
  Films}}},\ }\href {https://doi.org/10.1143/JJAP.38.5406} {\bibfield
  {journal} {\bibinfo  {journal} {Japanese Journal of Applied Physics}\
  }\textbf {\bibinfo {volume} {38}},\ \bibinfo {pages} {5406} (\bibinfo {year}
  {1999})}\BibitemShut {NoStop}%
\bibitem [{\citenamefont {Garcia}\ and\ \citenamefont
  {Bibes}(2014)}]{garcia_ferroelectric_2014}%
  \BibitemOpen
  \bibfield  {author} {\bibinfo {author} {\bibfnamefont {V.}~\bibnamefont
  {Garcia}}\ and\ \bibinfo {author} {\bibfnamefont {M.}~\bibnamefont {Bibes}},\
  }\bibfield  {title} {\bibinfo {title} {Ferroelectric tunnel junctions for
  information storage and processing},\ }\href
  {https://doi.org/10.1038/ncomms5289} {\bibfield  {journal} {\bibinfo
  {journal} {Nature Communications}\ }\textbf {\bibinfo {volume} {5}},\
  \bibinfo {pages} {4289} (\bibinfo {year} {2014})}\BibitemShut {NoStop}%
\bibitem [{\citenamefont {Chiu}(2014)}]{chiu_review_2014}%
  \BibitemOpen
  \bibfield  {author} {\bibinfo {author} {\bibfnamefont {F.-C.}\ \bibnamefont
  {Chiu}},\ }\bibfield  {title} {\bibinfo {title} {A {{Review}} on {{Conduction
  Mechanisms}} in {{Dielectric Films}}},\ }\href
  {https://doi.org/10.1155/2014/578168} {\bibfield  {journal} {\bibinfo
  {journal} {Advances in Materials Science and Engineering}\ }\textbf {\bibinfo
  {volume} {2014}},\ \bibinfo {pages} {e578168} (\bibinfo {year}
  {2014})}\BibitemShut {NoStop}%
\bibitem [{\citenamefont {Schr{\"o}der}\ \emph {et~al.}(2012)\citenamefont
  {Schr{\"o}der}, \citenamefont {Hau{\ss}mann}, \citenamefont {Thiessen},
  \citenamefont {Soergel}, \citenamefont {Woike},\ and\ \citenamefont
  {Eng}}]{schroder_conducting_2012}%
  \BibitemOpen
  \bibfield  {author} {\bibinfo {author} {\bibfnamefont {M.}~\bibnamefont
  {Schr{\"o}der}}, \bibinfo {author} {\bibfnamefont {A.}~\bibnamefont
  {Hau{\ss}mann}}, \bibinfo {author} {\bibfnamefont {A.}~\bibnamefont
  {Thiessen}}, \bibinfo {author} {\bibfnamefont {E.}~\bibnamefont {Soergel}},
  \bibinfo {author} {\bibfnamefont {T.}~\bibnamefont {Woike}},\ and\ \bibinfo
  {author} {\bibfnamefont {L.~M.}\ \bibnamefont {Eng}},\ }\bibfield  {title}
  {\bibinfo {title} {Conducting {{Domain Walls}} in {{Lithium Niobate Single
  Crystals}}},\ }\href {https://doi.org/10.1002/adfm.201201174} {\bibfield
  {journal} {\bibinfo  {journal} {Advanced Functional Materials}\ }\textbf
  {\bibinfo {volume} {22}},\ \bibinfo {pages} {3936} (\bibinfo {year}
  {2012})}\BibitemShut {NoStop}%
\bibitem [{\citenamefont {Schr{\"o}der}(2014)}]{schroder_conductive_2014}%
  \BibitemOpen
  \bibfield  {author} {\bibinfo {author} {\bibfnamefont {M.}~\bibnamefont
  {Schr{\"o}der}},\ }\emph {\bibinfo {title} {Conductive Domain Walls in
  Ferroelectric Bulk Single Crystals}},\ \href@noop {} {Ph.D. thesis},\
  \bibinfo  {school} {Technische Universit{\"a}t Dresden}, \bibinfo {address}
  {Dresden} (\bibinfo {year} {2014})\BibitemShut {NoStop}%
\bibitem [{\citenamefont {Ratzenberger}\ \emph {et~al.}(2024)\citenamefont
  {Ratzenberger}, \citenamefont {Kiseleva}, \citenamefont {Koppitz},
  \citenamefont {Beyreuther}, \citenamefont {Zahn}, \citenamefont {G{\"o}ssel},
  \citenamefont {Hegarty}, \citenamefont {Amber}, \citenamefont {R{\"u}sing},\
  and\ \citenamefont {Eng}}]{ratzenberger_reproducible_2024}%
  \BibitemOpen
  \bibfield  {author} {\bibinfo {author} {\bibfnamefont {J.}~\bibnamefont
  {Ratzenberger}}, \bibinfo {author} {\bibfnamefont {I.}~\bibnamefont
  {Kiseleva}}, \bibinfo {author} {\bibfnamefont {B.}~\bibnamefont {Koppitz}},
  \bibinfo {author} {\bibfnamefont {E.}~\bibnamefont {Beyreuther}}, \bibinfo
  {author} {\bibfnamefont {M.}~\bibnamefont {Zahn}}, \bibinfo {author}
  {\bibfnamefont {J.}~\bibnamefont {G{\"o}ssel}}, \bibinfo {author}
  {\bibfnamefont {P.~A.}\ \bibnamefont {Hegarty}}, \bibinfo {author}
  {\bibfnamefont {Z.~H.}\ \bibnamefont {Amber}}, \bibinfo {author}
  {\bibfnamefont {M.}~\bibnamefont {R{\"u}sing}},\ and\ \bibinfo {author}
  {\bibfnamefont {L.~M.}\ \bibnamefont {Eng}},\ }\bibfield  {title} {\bibinfo
  {title} {Toward the reproducible fabrication of conductive ferroelectric
  domain walls into lithium niobate bulk single crystals},\ }\href
  {https://doi.org/10.1063/5.0219300} {\bibfield  {journal} {\bibinfo
  {journal} {Journal of Applied Physics}\ }\textbf {\bibinfo {volume} {136}},\
  \bibinfo {pages} {104302} (\bibinfo {year} {2024})}\BibitemShut {NoStop}%
\bibitem [{\citenamefont {Zahn}\ \emph {et~al.}(2024)\citenamefont {Zahn},
  \citenamefont {Beyreuther}, \citenamefont {Kiseleva}, \citenamefont {Lotfy},
  \citenamefont {McCluskey}, \citenamefont {Maguire}, \citenamefont {Suna},
  \citenamefont {Rüsing}, \citenamefont {Gregg},\ and\ \citenamefont
  {Eng}}]{zahn_equivalentcircuit_2024}%
  \BibitemOpen
  \bibfield  {author} {\bibinfo {author} {\bibfnamefont {M.}~\bibnamefont
  {Zahn}}, \bibinfo {author} {\bibfnamefont {E.}~\bibnamefont {Beyreuther}},
  \bibinfo {author} {\bibfnamefont {I.}~\bibnamefont {Kiseleva}}, \bibinfo
  {author} {\bibfnamefont {A.~S.}\ \bibnamefont {Lotfy}}, \bibinfo {author}
  {\bibfnamefont {C.~J.}\ \bibnamefont {McCluskey}}, \bibinfo {author}
  {\bibfnamefont {J.~R.}\ \bibnamefont {Maguire}}, \bibinfo {author}
  {\bibfnamefont {A.}~\bibnamefont {Suna}}, \bibinfo {author} {\bibfnamefont
  {M.}~\bibnamefont {Rüsing}}, \bibinfo {author} {\bibfnamefont {J.~M.}\
  \bibnamefont {Gregg}},\ and\ \bibinfo {author} {\bibfnamefont {L.~M.}\
  \bibnamefont {Eng}},\ }\bibfield  {title} {\bibinfo {title}
  {Equivalent-circuit model that quantitatively describes domain-wall
  conductivity in ferroelectric \ce{LiNbO3}},\ }\href
  {https://doi.org/10.1103/PhysRevApplied.21.024007} {\bibfield  {journal}
  {\bibinfo  {journal} {Physical Review Applied}\ }\textbf {\bibinfo {volume}
  {21}},\ \bibinfo {pages} {024007} (\bibinfo {year} {2024})}\BibitemShut
  {NoStop}%
\bibitem [{\citenamefont {Kirbus}\ \emph {et~al.}(2019)\citenamefont {Kirbus},
  \citenamefont {Godau}, \citenamefont {Wehmeier}, \citenamefont {Beccard},
  \citenamefont {Beyreuther}, \citenamefont {Hau{\ss}mann},\ and\ \citenamefont
  {Eng}}]{kirbus_realtime_2019}%
  \BibitemOpen
  \bibfield  {author} {\bibinfo {author} {\bibfnamefont {B.}~\bibnamefont
  {Kirbus}}, \bibinfo {author} {\bibfnamefont {C.}~\bibnamefont {Godau}},
  \bibinfo {author} {\bibfnamefont {L.}~\bibnamefont {Wehmeier}}, \bibinfo
  {author} {\bibfnamefont {H.}~\bibnamefont {Beccard}}, \bibinfo {author}
  {\bibfnamefont {E.}~\bibnamefont {Beyreuther}}, \bibinfo {author}
  {\bibfnamefont {A.}~\bibnamefont {Hau{\ss}mann}},\ and\ \bibinfo {author}
  {\bibfnamefont {L.~M.}\ \bibnamefont {Eng}},\ }\bibfield  {title} {\bibinfo
  {title} {Real-{{Time 3D Imaging}} of {{Nanoscale Ferroelectric Domain Wall
  Dynamics}} in {{Lithium Niobate Single Crystals}} under {{Electric Stimuli}}:
  {{Implications}} for {{Domain-Wall-Based Nanoelectronic Devices}}},\ }\href
  {https://doi.org/10.1021/acsanm.9b01240} {\bibfield  {journal} {\bibinfo
  {journal} {ACS Applied Nano Materials}\ }\textbf {\bibinfo {volume} {2}},\
  \bibinfo {pages} {5787} (\bibinfo {year} {2019})}\BibitemShut {NoStop}%
\bibitem [{\citenamefont {Godau}\ \emph {et~al.}(2017)\citenamefont {Godau},
  \citenamefont {K{\"a}mpfe}, \citenamefont {Thiessen}, \citenamefont {Eng},\
  and\ \citenamefont {Hau{\ss}mann}}]{godau_enhancing_2017}%
  \BibitemOpen
  \bibfield  {author} {\bibinfo {author} {\bibfnamefont {C.}~\bibnamefont
  {Godau}}, \bibinfo {author} {\bibfnamefont {T.}~\bibnamefont {K{\"a}mpfe}},
  \bibinfo {author} {\bibfnamefont {A.}~\bibnamefont {Thiessen}}, \bibinfo
  {author} {\bibfnamefont {L.~M.}\ \bibnamefont {Eng}},\ and\ \bibinfo {author}
  {\bibfnamefont {A.}~\bibnamefont {Hau{\ss}mann}},\ }\bibfield  {title}
  {\bibinfo {title} {Enhancing the {{Domain Wall Conductivity}} in {{Lithium
  Niobate Single Crystals}}},\ }\href {https://doi.org/10.1021/acsnano.7b01199}
  {\bibfield  {journal} {\bibinfo  {journal} {ACS Nano}\ }\textbf {\bibinfo
  {volume} {11}},\ \bibinfo {pages} {4816} (\bibinfo {year}
  {2017})}\BibitemShut {NoStop}%
\bibitem [{\citenamefont {Guyonnet}\ \emph {et~al.}(2011)\citenamefont
  {Guyonnet}, \citenamefont {Gaponenko}, \citenamefont {Gariglio},\ and\
  \citenamefont {Paruch}}]{Guyonnet_Conduction_2011}%
  \BibitemOpen
  \bibfield  {author} {\bibinfo {author} {\bibfnamefont {J.}~\bibnamefont
  {Guyonnet}}, \bibinfo {author} {\bibfnamefont {I.}~\bibnamefont {Gaponenko}},
  \bibinfo {author} {\bibfnamefont {S.}~\bibnamefont {Gariglio}},\ and\
  \bibinfo {author} {\bibfnamefont {P.}~\bibnamefont {Paruch}},\ }\bibfield
  {title} {\bibinfo {title} {{Conduction at Domain Walls in Insulating
  Pb(Zr$_{0.2}$Ti$_{0.8}$)O$_3$ Thin Films}},\ }\href
  {https://doi.org/10.1002/adma.201102254} {\bibfield  {journal} {\bibinfo
  {journal} {Advanced Materials}\ }\textbf {\bibinfo {volume} {23}},\ \bibinfo
  {pages} {5377} (\bibinfo {year} {2011})}\BibitemShut {NoStop}%
\bibitem [{\citenamefont {Godau}(2018)}]{godau_herstellung_2018}%
  \BibitemOpen
  \bibfield  {author} {\bibinfo {author} {\bibfnamefont {C.}~\bibnamefont
  {Godau}},\ }\emph {\bibinfo {title} {{Herstellung und Charakterisierung
  hochleitf{\"a}higer, ferroelektrischer Dom{\"a}nenw{\"a}nde}}},\ \href@noop
  {} {Ph.D. thesis},\ \bibinfo  {school} {Technische Universit{\"a}t Dresden},
  \bibinfo {address} {Dresden} (\bibinfo {year} {2018})\BibitemShut {NoStop}%
\bibitem [{sup()}]{supplement}%
  \BibitemOpen
  \href@noop {} {}\bibinfo {note} {See Supplemental Material at [final URL will
  be inserted by publisher] for further details on the mathematical background,
  the measurement setup, further reference data recorded on bulk lithium
  niobate and a commercial Schottky diode, additional electrical transport
  data, as well as extensive I-V-curve fit data, which includes Refs. [37],
  [38], [39].}\BibitemShut {Stop}%
\bibitem [{\citenamefont {Zahn}(2022)}]{zahn_nonlinear_2022}%
  \BibitemOpen
  \bibfield  {author} {\bibinfo {author} {\bibfnamefont {M.}~\bibnamefont
  {Zahn}},\ }\emph {\bibinfo {title} {Nonlinear Electronic Conductivity in
  Lithium Niobate Domain Walls}},\ \href@noop {} {Master's thesis},\ \bibinfo
  {school} {Technische Universit{\"a}t Dresden}, \bibinfo {address} {Dresden}
  (\bibinfo {year} {2022})\BibitemShut {NoStop}%
\bibitem [{\citenamefont {Meyer}\ \emph {et~al.}(2014)\citenamefont {Meyer},
  \citenamefont {Nataf},\ and\ \citenamefont {Granzow}}]{meyer_field_2014}%
  \BibitemOpen
  \bibfield  {author} {\bibinfo {author} {\bibfnamefont {N.}~\bibnamefont
  {Meyer}}, \bibinfo {author} {\bibfnamefont {G.~F.}\ \bibnamefont {Nataf}},\
  and\ \bibinfo {author} {\bibfnamefont {T.}~\bibnamefont {Granzow}},\
  }\bibfield  {title} {\bibinfo {title} {Field induced modification of defect
  complexes in magnesium-doped lithium niobate},\ }\href
  {https://doi.org/10.1063/1.4905021} {\bibfield  {journal} {\bibinfo
  {journal} {Journal of Applied Physics}\ }\textbf {\bibinfo {volume} {116}},\
  \bibinfo {pages} {244102} (\bibinfo {year} {2014})}\BibitemShut {NoStop}%
\bibitem [{\citenamefont {Du}\ \emph {et~al.}(2025)\citenamefont {Du},
  \citenamefont {Drozdov}, \citenamefont {Minkov}, \citenamefont {Balakirev},
  \citenamefont {Kong}, \citenamefont {Smith}, \citenamefont {Yan},
  \citenamefont {Shen}, \citenamefont {Gegenwart},\ and\ \citenamefont
  {Eremets}}]{du_superconducting_2025}%
  \BibitemOpen
  \bibfield  {author} {\bibinfo {author} {\bibfnamefont {F.}~\bibnamefont
  {Du}}, \bibinfo {author} {\bibfnamefont {A.~P.}\ \bibnamefont {Drozdov}},
  \bibinfo {author} {\bibfnamefont {V.~S.}\ \bibnamefont {Minkov}}, \bibinfo
  {author} {\bibfnamefont {F.~F.}\ \bibnamefont {Balakirev}}, \bibinfo {author}
  {\bibfnamefont {P.}~\bibnamefont {Kong}}, \bibinfo {author} {\bibfnamefont
  {G.~A.}\ \bibnamefont {Smith}}, \bibinfo {author} {\bibfnamefont
  {J.}~\bibnamefont {Yan}}, \bibinfo {author} {\bibfnamefont {B.}~\bibnamefont
  {Shen}}, \bibinfo {author} {\bibfnamefont {P.}~\bibnamefont {Gegenwart}},\
  and\ \bibinfo {author} {\bibfnamefont {M.~I.}\ \bibnamefont {Eremets}},\
  }\bibfield  {title} {\bibinfo {title} {Superconducting gap of {{H3S}}
  measured by tunnelling spectroscopy},\ }\href
  {https://doi.org/10.1038/s41586-025-08895-2} {\bibfield  {journal} {\bibinfo
  {journal} {Nature}\ }\textbf {\bibinfo {volume} {641}},\ \bibinfo {pages}
  {619} (\bibinfo {year} {2025})}\BibitemShut {NoStop}%
\bibitem [{\citenamefont {Kim}\ \emph {et~al.}(2011)\citenamefont {Kim},
  \citenamefont {Song}, \citenamefont {Strigl}, \citenamefont {Pernau},
  \citenamefont {Lee},\ and\ \citenamefont {Scheer}}]{kim_conductance_2011}%
  \BibitemOpen
  \bibfield  {author} {\bibinfo {author} {\bibfnamefont {Y.}~\bibnamefont
  {Kim}}, \bibinfo {author} {\bibfnamefont {H.}~\bibnamefont {Song}}, \bibinfo
  {author} {\bibfnamefont {F.}~\bibnamefont {Strigl}}, \bibinfo {author}
  {\bibfnamefont {H.-F.}\ \bibnamefont {Pernau}}, \bibinfo {author}
  {\bibfnamefont {T.}~\bibnamefont {Lee}},\ and\ \bibinfo {author}
  {\bibfnamefont {E.}~\bibnamefont {Scheer}},\ }\bibfield  {title} {\bibinfo
  {title} {Conductance and {{Vibrational States}} of {{Single-Molecule
  Junctions Controlled}} by {{Mechanical Stretching}} and {{Material
  Variation}}},\ }\href {https://doi.org/10.1103/PhysRevLett.106.196804}
  {\bibfield  {journal} {\bibinfo  {journal} {Physical Review Letters}\
  }\textbf {\bibinfo {volume} {106}},\ \bibinfo {pages} {196804} (\bibinfo
  {year} {2011})}\BibitemShut {NoStop}%
\end{thebibliography}

%

\clearpage

\setcounter{table}{0}
\renewcommand{\thetable}{S\arabic{table}}  

\setcounter{page}{1}
\renewcommand{\thepage}{S\arabic{page}}

\setcounter{equation}{0}
\renewcommand{\theequation}{S.\arabic{equation}} 

\setcounter{figure}{0}
\renewcommand{\thefigure}{S\arabic{figure}}

\renewcommand{\thesection}{\Alph{section}}

\renewcommand{\thesubsection}{\arabic{subsection}}

\onecolumngrid

\titleformat{\section}{\centering\normalfont\bfseries}{\thesection.}{.3em}{\vspace{.3ex}}
\titleformat{\subsection}{\centering\normalfont\bfseries}{\thesection.\thesubsection}{.3em}{\vspace{.3ex}}

\part{Supplemental Material}

\section{Reference data I: \texorpdfstring{\ce{LiNbO3}}{LiNbO3} bulk conductivity}
\label{sec:app:bulkac}

\begin{figure}[ht]
    \centering
    \includegraphics[width=15cm]{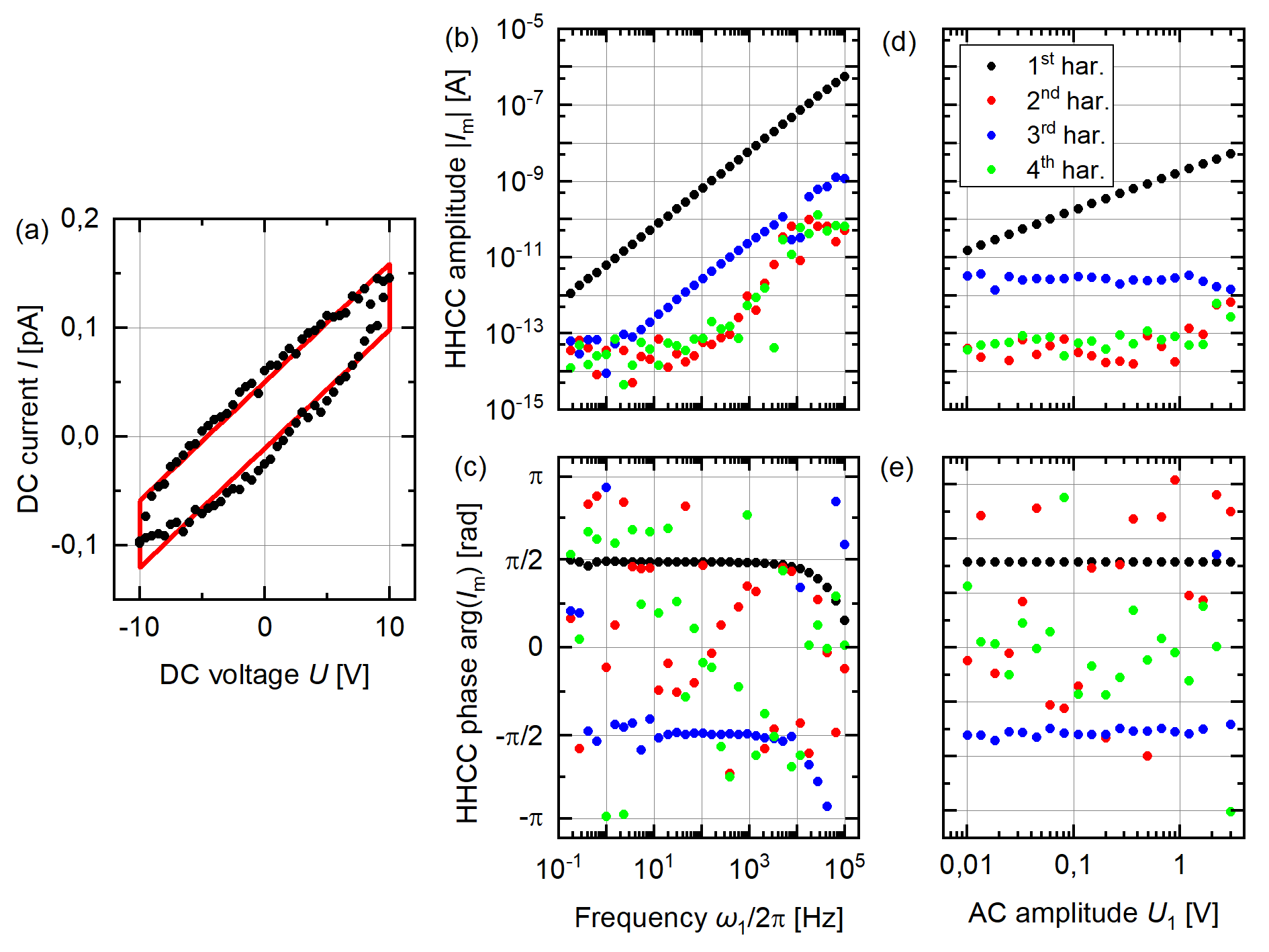}
    \vspace{2mm}
    {{\phantomsubfloat{\label{fig:app:bulk_dc}}
     \phantomsubfloat{\label{fig:app:bulk_ac_frequency_amplitude}}
     \phantomsubfloat{\label{fig:app:bulk_ac_frequency_phase}}
     \phantomsubfloat{\label{fig:app:bulk_ac_excit_amp_amplitude}}
     \phantomsubfloat{\label{fig:app:bulk_ac_excit_amp_phase}}}
    \caption{Electric conductivity of mono-domain \ce{LiNbO3} with evaporated chromium electrodes: \textbf{(a)} DC conductivity, showing next to the resistive part a hysteresis due to capacitive charging by the external electrodes, $\text{d}U/\text{dt} = \SI{0.26}{V/s}$; \textbf{(b,c)} AC conductivity with respect to the fundamental frequency $\omega_1/2 \pi$ (constant excitation parameters: $U_1 = \SI{300}{mV}$, $U_0 = \SI{0}{V}$) and \textbf{(d,e)} with respect to the AC excitation amplitude $U_1$ (constant excitation parameters: $\omega_1/2\pi = \SI{84}{Hz}$, $U_0 = \SI{0}{V}$).}}
    \label{fig:app:bulk_conductivity}
\end{figure}

To countercheck whether the DC and AC current-voltage data acquired on the domain wall samples \emph{DW-1} and \emph{DW-2} really probes the conductance of the DWs and not of the bulk, reference measurements of the DC and AC current were accomplished on a piece of monodomain \ce{LiNbO3} of same geometry as the two samples with artificially grown domain walls. The results are shown in \cref{fig:app:bulk_conductivity}.

Figure~\ref{fig:app:bulk_dc} addresses the DC conductivity that can be understood as a parallel circuit of a resistor and a capacitor. Assuming a triangular-shaped applied voltage with respect to time, a parallelogram-like I-V curve is expected and observed as shown by the red solid line. Thereby the slope of the diagonal parts is determined by the resistor (approx.~\SI{90}{\tera\ohm}). The corresponding currents are seven orders of magnitude lower as compared to the currents observed for the conductivity-enhanced domain walls (cf.~\cref{fig:lno_iu_curves} of the main text) and confirm the striking contribution of the latter to the total conductance. Furthermore, the area of the hysteresis loop (and its height given by $2 \, I_\text{C}$, with $I_\text{C}$ being the charging current) are determined by the capacity $C$ and voltage-sweep velocity $\frac{\text{d}U}{\text{d}t}$ via the fundamental law for the charging current of capacitors: \(I_\text{C} = C \frac{\text{d}U}{\text{d}t}\). A rough estimation, using a voltage sweep velocity of $\text{d}U/\text{d}t = \SI{0.26}{s}$, a relative dielectric constant of $\varepsilon_\text{r} = \num{33}$ (see ref.~\cite{meyer_field_2014}), and an electrode area of $A = \SI{0.25}{mm^2}$, predicts $I_\text{C} = \SI{91}{fA}$, which is of the same order of magnitude as the experimentally observed value \(I_\text{C, exp.} = \SI{30.6}{fA}\). While the resistive and capacitive current are of the same order of magnitude in the DC bulk measurement, the situation is fundamentally different in samples with domain walls. Since there the resistive current increases due to the (conductive) domain walls by six orders magnitude, while the capacitive current increases by two orders of magnitude only in the AC measurements (due to the same increase in frequency compared to "DC" case), the capacitive current is relatively low, as assumed over the entire analysis.

\Cref{fig:app:bulk_ac_frequency_amplitude,fig:app:bulk_ac_frequency_phase} visualize the frequency dependence of the HHCCs (that is not discussed in the main text). Amplitude and phase of the first harmonic order fully agree with a purely capacitive contribution. A phase rotation is observed beyond \SI{e4}{\hertz} due to the near current-amplifier bandwidth. Further, the second and fourth harmonic orders' behavior are easy to explain, since they are consequently below both the absolute and relative noise limits regarding the amplitude and exhibit a randomly distributed phase. The systematic increase of the third harmonic is most likely due the linear propagation of the parasitic third-harmonic contribution produced by the signal generator.

The analysis is completed by the excitation amplitude dependence shown in \cref{fig:app:bulk_ac_excit_amp_amplitude,fig:app:bulk_ac_excit_amp_phase}. In agreement with the previously drawn picture of a dominant capacitive behavior, the amplitude of the first-harmonic current contribution increases linearly with the excitation amplitude while the phase is constant at $+\pi/2$. The odd harmonics two and four are below the noise limit and thus the corresponding phase angle is not reliably measurable. Only the third harmonic reaches a significant level and stays constant as a function of the excitation amplitude -- a behavior that can be traced back to the digitization error within the direct digital synthesis (DDS) of the excitation sine wave within the signal generator.

In summary, the monodomain lithium niobate crystals acts as a close-to-ideal capacitor. This is significantly different -- qualitatively and quantitatively regarding the order of magnitude for the HHCC amplitudes -- from the results shown in the main text for the samples containing domain walls.

\section{Mathematical background}
\label{sec:app:math_background}


The following considerations extend the mathematical introduction given in \cref{sec:methods:math} by deriving the general relation between the DC I-V curve of a passive one-port circuit element (containing two contacts and no internal voltage source) and the Fourier coefficients of the electric current under sinusoidal excitation. As said in the main text, these Fourier coefficients are called higher-harmonic current contribution (HHCCs) within our framework. 
As also said before, we assume the absence of capacitive or inductive components, so the the electric current is fully determined by the DC current-voltage characteristic, formally expressed as:

\begin{equation*}
	\forall t: I(t) = I_\text{DC}(U(t)).
\end{equation*}

In most cases, this condition is fulfilled in the low-frequency limit (e.g., within simple capacitive contributions the capacitive current scales with $1/\omega$) and motivates our chosen frequency range of \SI{10}{\hertz} to \SI{1}{\kilo\hertz} that is also a compromise with respect to the total measurement time. Starting from an arbitrary I-V curve that is Taylor-expandable around a constant offset voltage $U_0$,

\begin{equation}
    I_\text{DC}(U) = I(U_0 + \Delta U) = \sum_{k = 0}^\infty a_k \cdot \Delta U^k, \quad
    a_k = \left. \frac{\text{d}^k I_\text{DC}}{\text{d} U^k} \right|_{U = U_0}
    \label{equ:app_math:taylor_expansion}
\end{equation}

we aim to derive the HHCCs $I_m$ (as a function the of Taylor coeffients $a_k$) that are given by (replication of \cref{equ:fourier_integral}):

\begin{equation*}
    I_m = i \int_0^{2 \pi/\omega_1}  I_\text{DC}(U_0 + U_1 \sin(i \omega_1 t)) \, \exp(- i m \omega_1 t) \text{d}t.
\end{equation*}

To solve the integral, in a first step the expansion introduced in \cref{equ:app_math:taylor_expansion} is plugged into the expression and the sinusiodial excitation $U_1 \sin(\omega_1 t)$ takes the role of the variation $\Delta U$. It is further convenient to calculate real and imaginary part separately while the procedure is the same for both parts and we focus here on the real part that is given by

\begin{equation}\begin{aligned}
	\Re(I_m) &= \int_0^{2\pi/\omega_1} I_\text{DC}(U(t)) \sin(m \omega_1 t) \text{d} t
        = \int_0^{2\pi/\omega_1} \left[
        \sum_{k = 0}^\infty a_k (U_1 \sin(\omega_1 t))^k
        \right] \sin(m \omega_1 t) \text{d} t \\
    &= \sum_{k = 0}^\infty a_k U_1^k \int_0^{2 \pi/\omega_1}
        \sin(m \omega_1 t) \sin^k(\omega_1 t) \text{d}t.
	\label{equ:app_math:real_fourier_integral}
\end{aligned}\end{equation}

The remaining integral can be solved as shown in more detail in ref.~\cite{zahn_nonlinear_2022} (equ.~C.4 to C.8) by expanding the power of sine (introducing a binomial coefficient) and by the reverse application of the trigonometric addition theorem -- and finally results in:

\begin{equation}
    \Re(I_m) = \sin\left(-m \frac{\pi}{2}\right) \sum_{l = 0}^\infty a_{2l + m}
        \left( \frac{U_1}{2} \right)^{2l + m} \binom{2l + m}{l}.
\end{equation}

For the imaginary part, the result is same except of the first factor that becomes $\cos\left(m \frac{\pi}{2}\right)$ -- so the final expression is the following:

\begin{equation}\begin{aligned}
    I_m &= \sum_{l = 0}^\infty a_{2l + m} \left( \frac{U_1}{2} \right)^{2l + m} \binom{2l + m}{l}
        \left[ \sin\left(-m \frac{\pi}{2}\right) + i \cos\left(m \frac{\pi}{2}\right) \right] \\
        &= \sum_{l = 0}^\infty a_{2l + m} \left( \frac{U_1}{2} \right)^{2l + m} \binom{2l + m}{l}
        \exp\left(- i \frac{m - 1}{2} \pi \right).
    \end{aligned}
   \label{equ:app_math:hhcc_general}
\end{equation}

The phase factor changes by $\SI{90}{\degree} \hat{=} \pi/2$ between neighboring harmonic orders $m$ that is discussed in the main text as observation (II) in \cref{sec:results:consistency}. Further, the contributions to the sum can cancel out each other, i.e. by different signs of the coefficients $a_{2 l + m}$ and $I_m$, and thus can become zero as discussed in \cref{sec:results:consistency} as observation (III).


The general expression for the HHCCs in \cref{equ:app_math:hhcc_general} can be easily applied for transport models of power law type as ohmic transport or space charged limited conduction (SCLC) (see \cref{tab:conduction_mechanisms}). Another case with an analytical solution is the unidirectional hopping transport that is close to the experimentally observed behavior of \textit{DW-2} (see \cref{fig:lno_dc_dw2}).

In order to make use of the above general expression, we have to determine the Taylor coefficients $a_k$ first. In the present case, this can be achieved by applying the Taylor expansion of the exponential function on the Shockley equation as listed in \cref{tab:conduction_mechanisms}, reordering the term, and compare it to \cref{equ:app_math:taylor_expansion}, resulting in:

\begin{equation}\begin{aligned}
    I(U) &= I_\text{HT} \left[ \exp \left(\frac{U}{U_\text{HT}}\right) - 1 \right] 
    = I_\text{HT} \left[ \exp \left(\frac{U_0 + \Delta U}{U_\text{HT}}\right) - 1 \right] \\
    &= I_\text{HT} \exp\left(\frac{U_0}{U_\text{HT}}\right)
    \exp\left(\frac{\Delta U}{U_\text{HT}}\right) - I_\text{HT} \\
    &= I_\text{HT} \exp\left(\frac{U_0}{U_\text{HT}}\right) \left[\sum_{k = 0}^\infty
    \frac{1}{k!} \left(\frac{\Delta U}{U_\text{HT}}\right)^k \right] - I_\text{HT} \\
    &= - I_\text{HT} + \sum_{k = 0}^\infty \underbrace{\frac{I_\text{HT}}{k! \ U_\text{HT}^k} \exp\left(\frac{U_0}{U_\text{HT}}\right) }_{:= a_{k, \text{HT}}} \Delta U^k,
    \end{aligned}
    \label{equ:app_math:shockley_taylor_coefficients}
\end{equation}
and $a_{0, \text{HT}} = I_\text{HT} \left(\exp(U_0 / U_\text{HT}) - 1\right)$. Inserting the result of \cref{equ:app_math:shockley_taylor_coefficients} into the general prediction of \cref{equ:app_math:hhcc_general} reveals the expected HHCCs of a single Schottky diode, which then can be expressed in a compact form by reminding the definition of the modified Bessel function of first kind:

\begin{equation*}
    \mathcal{I}_m (x) = \sum_{l = 0}^\infty \frac{(x/2)^{2l + m}}{(m + l)! \, l!}.
\end{equation*}

The "compact" form is given by:

\begin{equation*}
   I_m = I_\text{HT} \cdot
       \mathcal{I}_m \left( \frac{U_1}{U_\text{HT}} \right)
       \exp \left( \frac{U_0}{U_\text{HT}} \right) \exp \left(
       - i \frac{m - 1}{2} \pi \right),
\end{equation*}
as stated in the main text as \cref{equ:diode_fourier_coefficients}.

In the last part of this chapter we discuss some further properties of the HHCCs that play a role in the main text or are important for the experimental realization of HHCC measurements. 

First, \textit{single} higher harmonics of the electric current have been commonly used to detect features within a particular derivative of the I-V curve in the past, for example, the gap width in superconductors \cite{du_superconducting_2025} via the first harmonic or the quantized conduction in a single molecule via the second harmonic \cite{kim_conductance_2011}. This approach can be validated as it is a special case of \cref{equ:app_math:hhcc_general}. The key requirement therefore is a \emph{weak non-linearity} as defined in the following: If the Taylor coefficients in \cref{equ:app_math:taylor_expansion} decay strongly, which means that each $a_k$ is significantly smaller than the preceding order $a_{k-1}$, then every term in the sums of \cref{equ:app_math:taylor_expansion} and \cref{equ:app_math:hhcc_general} contributes much weaker than the previous one. Accordingly, the sum in \cref{equ:app_math:hhcc_general} is dominated by the first term with $l = 0$ and all following terms can be neglected. In turn, the contribution $I_m$ becomes proportional to the Taylor coefficient $a_m$ that represents -- according to the initial definition in \cref{equ:app_math:taylor_expansion} -- the $m^\text{th}$ derivative of the I-V~curve. Consequently, in total we can write:

\begin{equation}\begin{aligned}
    \text{\cref{equ:app_math:taylor_expansion}:}
    && \frac{\text{d}^m I_\text{DC}(U)}{\text{d} U^m}
    & = \sum_{k = 0}^{\infty} a_{k + m} \frac{(k + m) !}{k!} \Delta U^k & & \underset{\Delta U = 0}{=} a_m \cdot m! \\
    \text{\cref{equ:app_math:hhcc_general}:}
    && I_m &\underset{l = 0}{=} 
		a_{m} \left( \frac{U_1}{2} \right)^{m}
		\binom{m}{0} \exp\left(- i \frac{m - 1}{2} \pi \right)
        & & \underset{\hphantom{\Delta U = 0}}{=} a_{m}
            \cdot \left( \frac{U_1}{2} \right)^{m} (-i)^{m - 1}.
\end{aligned}\end{equation}

This proves:

\begin{equation*}
    I_m \propto a_m \propto \frac{\text{d}^m I_\text{DC}(U)}{\text{d}U^m},
\end{equation*}
meaning that $I_m$ is indeed a measure of the $m^\text{th}$ derivative of the static I-V ~curve and this has exactly been observed in the domain wall samples of the present study (cf. observation (I) in \cref{sec:results:consistency}).

A second annotation is directly related to the HHCC measuring process: commonly available lock-in amplifiers extract the root-mean square of the input signal, so the  experimentally measured coefficients are a factor of \(\sqrt{2} \pi\) smaller then expected from the evaluation of \cref{equ:app_math:hhcc_general}, respectively \cref{equ:diode_fourier_coefficients}.

Finally, despite it is mathematically sufficient to integrate over a single period (\cref{equ:fourier_integral}), it is experimentally required to average over several periods for improved noise suppression. In this conjunction, the determining parameter is the \emph{integration time} $\tau$ of the lock-in amplifier that has been chosen in all experiments between 10 and 100 times the excitation frequency - again as a trade off with respect to the total measurement time.

\section{Details on the experimental realization of the HHCC analysis}
\label{sec:app:experimental_realization}

\begin{figure*}[htb]
	\centering
	\includegraphics[width=12cm]{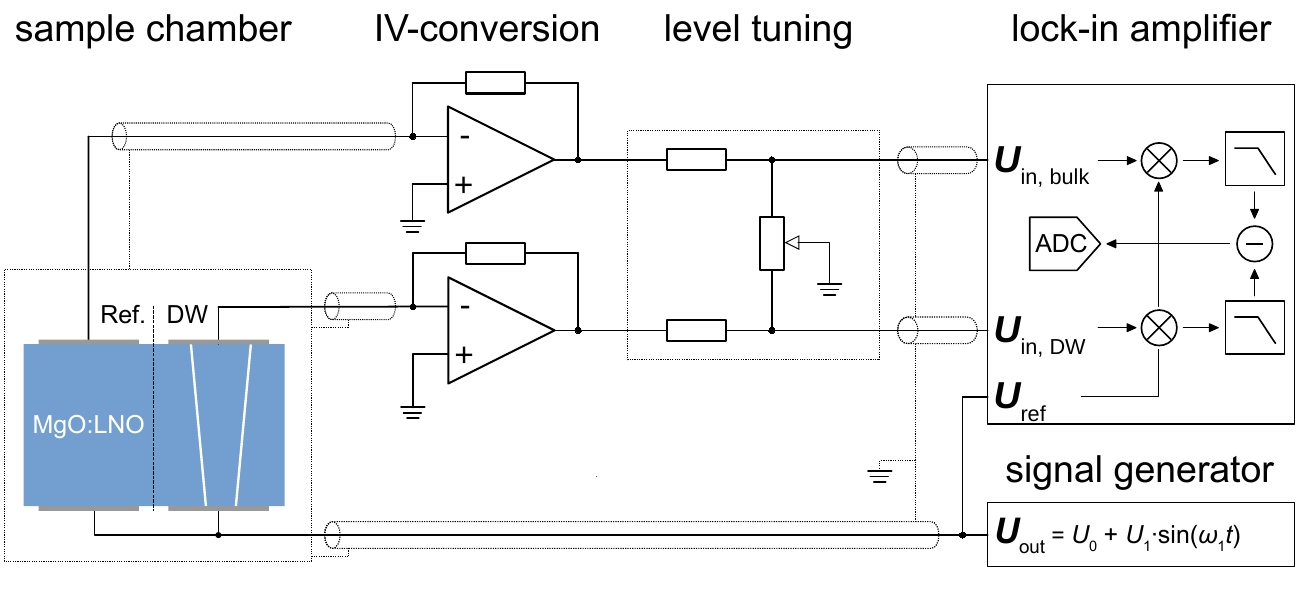}
	\caption{Detailed sketch of the experimental setup applied to record Fourier coefficients (HHCCs) of the electric current under AC voltage excitation. Signal generation, I-V conversion, and lock-in detection were realized with commercial instruments, while sample chamber and level tuning were home-built devices. To compensate the parasitic capacitor formed by the metal contact electrodes with the sample bulk material around the DWs of interest, a second pair of electrodes was deposited on the neighboring purely monodomain bulk material. Thus, the setup was completed towards a fully differential measurement. The used instruments are specified in \cref{tab:setup_components}.}
	\label{fig:experimental_setup}
\end{figure*}

A more detailed version of the experimental setup, extending the sketch of \cref{fig:ac_principle:circuit} of the main text, is given in \cref{fig:experimental_setup}. It includes two additional stages required to ensure the functionality. 

First, the current-to-voltage conversion has to be addressed. Due to the typically low electric currents, measuring the voltage across a shunt resistor is not possible and an additional circuit, in this case measuring the current against a virtual ground, is required. The currents within the signal and reference paths are measured separately, as there is a second step of post-processing required. 

Second, as shown also in the figure, the area of sample and reference electrodes always slightly differ in size leading (according to \cref{sec:app:bulkac}) to different capacitive contributions. This "artifact" is compensated by the level tuning circuit depicted in \cref{fig:experimental_setup}. Using the high-precision trimming potentiometer, the first-order capacitive contributions at low frequency and without an additional offset voltage are equalized, as they are purely arising from the plate capacitors formed by the electrodes.

The devices, which were used to realize the experimental setup in \cref{fig:experimental_setup} are listed in \cref{tab:setup_components}. Two different device combinations -- each of them providing specific advantages -- were employed: (i) The "fast-acquisition" setup profits from the different demodulators within the \textit{Zurich Instruments UHFLI} lock-in amplifier, which enable to measure up to 8 different harmonic orders at the same time, reducing the measurement time significantly. This setup was used for measurements on sample \textit{DW-1}, where the conductivity was sufficiently high and the current resolution limit not the limiting factor. (ii) In contrast, a "high-precision" setup was required to deal with the comparably low currents observed in sample \textit{DW-2}. Thereby the high-quality input stage and the availability of longer time constants of the \textit{Stanford Research SR830} lock-in amplifier pushed the resolution limit, while all harmonic orders had to be measured sequentially.

\begin{table}[htb]
	{\renewcommand{\arraystretch}{1.3}
    \begin{tabular}{c|c|c} \hline \hline
		\thead{component} & \thead{"fast-acquisition" setup} & \thead{"high-precision"setup} \\ \hline \hline
		signal generator & \gape{\makecell{\textit{ZI UHFLI} LIA internal \\ generator + \textit{ITACO} \\ \textit{4302} preamplifier}}
			& \textit{Agilent 33250A} \\
		sample chamber & \multicolumn{2}{c}{home-built} \\
		I-V converter & \multicolumn{2}{c}{\textit{Femto DLPCA-200}} \\
		level tuning & \multicolumn{2}{c}{home-built} \\
		LIA & \textit{ZI UHLFI} & \textit{SR 830} \\ \hline
        Sample & \textit{DW-1} & \textit{DW-2} \\ \hline \hline
	\end{tabular}}
	\caption{Measurement instruments used to implement a "fast-acquisition" and "high-precision" version of the experimental setup principle shown in \cref{fig:experimental_setup}.}
	\label{tab:setup_components}
\end{table}

\newpage

\section{Reference data II: Higher-harmonic current contributions of a commercial Schottky diode}
\label{sec:app:bat48}

\begin{figure*}[htb]
	\includegraphics[width=16.5cm]{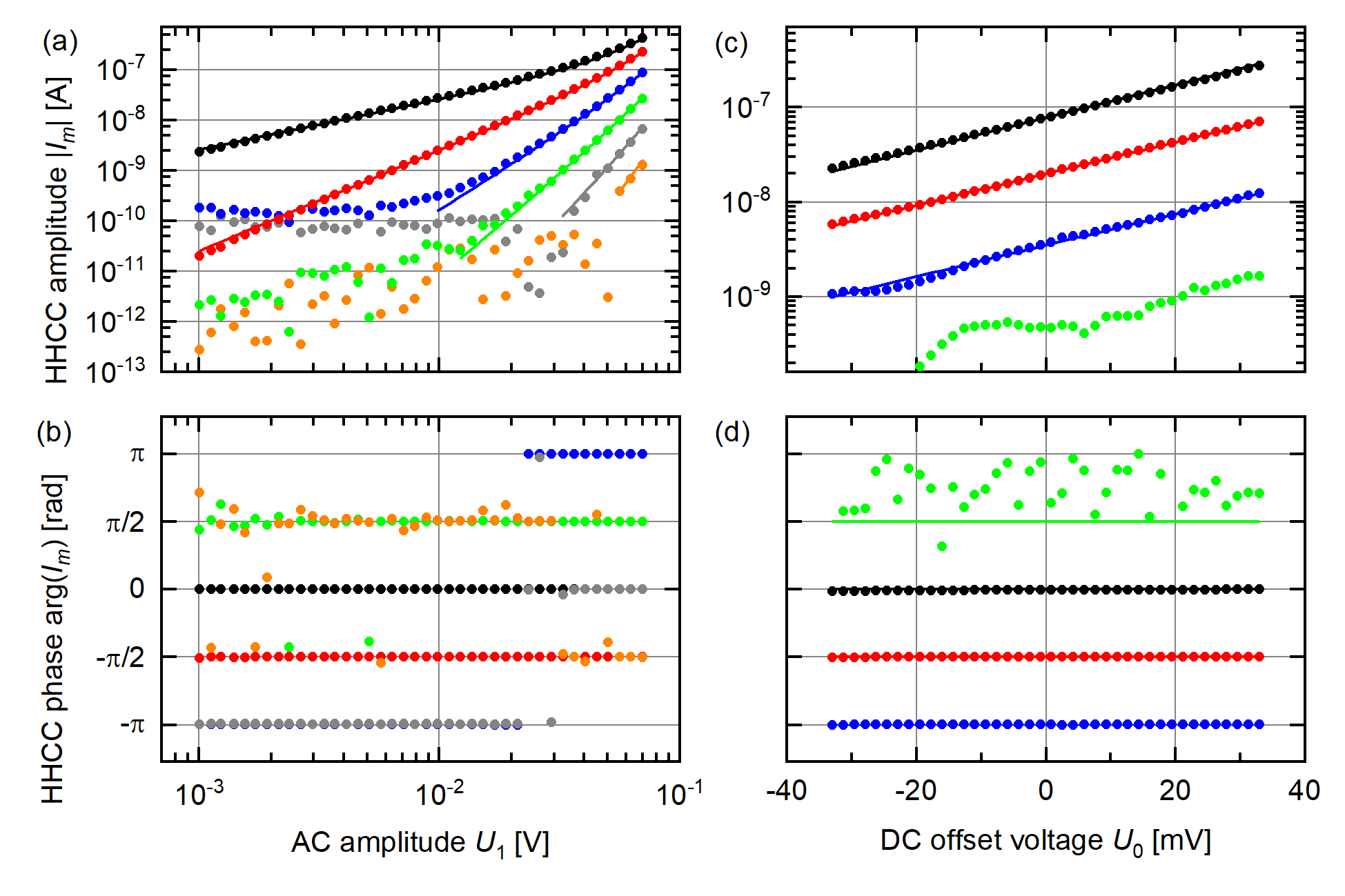}
	{\phantomsubfloat{\label{fig:bat48:ampsweep:amp}}
     \phantomsubfloat{\label{fig:bat48:ampsweep:phase}}
     \phantomsubfloat{\label{fig:bat48:offsetsweep:amp}}
     \phantomsubfloat{\label{fig:bat48:offsetsweep:phase}}}
	\caption{Higher-harmonic current contributions of a commercial Schottky diode of type \textit{BAT 48}. \textbf{(a)} Absolute value and \textbf{(b)} phase of the electric current's Fourier coefficients \(I_m\) with respect to the AC excitation amplitude \(U_1\). Solid lines represent the best fitting \emph{modified Bessel function of first kind} according to \cref{equ:diode_fourier_coefficients}. The color coding is the same as in \cref{fig:ac_principle:complex_plane}. \(f = \SI{86}{\hertz}\), \(U_0 = 0\), \(\tau = \SI{0.94}{s}\). \textbf{(c)} Amplitude and \textbf{(d)} phase of the Fourier coefficients \(I_m\) measured as a function of the DC offset voltage $U_0$. \(f = \SI{270}{\hertz}\), \(U_1 = \SI{20}{\milli\volt_{RMS}}\), \(\tau = \SI{0.27}{s}\), measured by \textit{fast-acquisition} setup.}
	\label{fig:bat48-hhcc}
\end{figure*}

Before starting the measurements on the \ce{LiNbO3} samples, the HHCC analysis was tested on a commercial Schottky diode of type \textit{BAT48} (\textit{STMicroelectronics}, \textit{DO-213AA} package), since for the specific circuit element of a single diode there is an analytic prediction of the HHCCs available as discussed in \cref{sec:methods:math} and summarized in \cref{equ:diode_fourier_coefficients}. The experimental results are shown in \cref{fig:bat48-hhcc}, including both the amplitude and offset voltage dependence of HHCC amplitudes and phases. To stay within the same current range as compared to the \ce{LiNbO3} samples, significantly lower voltages were applied due to the lower resistance of the commercial diode.

The HHCC amplitude with respect to the AC excitation amplitude $U_1$ is shown in \cref{fig:bat48:ampsweep:amp}. Compared to the solid lines, representing the best fitting modified Bessel functions, the experimental data (dots) match the theoretical predictions well as long as the currents are above the detection limit of around \SI{3e-10}{A}. From the theoretical prediction in \cref{equ:diode_fourier_coefficients}, a change of slope within the amplitude in the $\log_{10}\abs{I_m}$-vs.-$U_1$ dependence is expected at the characteristic voltage $U_\text{HT} = n k_\text{B} T / q$ (with $q$ being the elementary charge and $n$ the ideality factor) that is indeed slightly visible around $U_1 = \SI{35}{mV}$. By a joint fitting process of all measured data sets up to the sixth harmonic order, the diode parameters, i.e., the saturation current $I_\text{HT}$ and characteristic voltage $U_\text{HT}$, were evaluated to be \(I_\text{HT} = \SI{1.3 \pm 0.6e-7}{A}\) and $U_\text{HT} = \SI{25.5 \pm 3.8}{mV}$, the latter indicating a reasonable ideality factor $n$ around $1.0$, as expected for a conventional silicon-based diode.

Moving on to the HHCC phase displayed in \cref{fig:bat48:ampsweep:phase}, a strict anti-clockwise rotation is observed as predicted from \cref{equ:app_math:hhcc_general} over the full amplitude range. This holds as well for the phase of the offset voltage dependence shown in \cref{fig:bat48:offsetsweep:phase}.

Finally, the HHCC amplitude as a function of the offset voltage $U_0$ is shown in \cref{fig:bat48:offsetsweep:amp}. In agreement with \cref{equ:diode_fourier_coefficients} an exponential dependence is observed and the characteristic voltage can be evaluated again, using a joint fitting process of the acquired four harmonic orders, turning out to be $U_\text{HT} = \SI{26.24 \pm 0.02}{meV}$, which also corresponds to a reasonable ideality factor of $n = \num{1.02}$.

In conclusion, this test experiment confirmed all predictions from the analytical calculation based on \cref{equ:diode_fourier_coefficients} and thus validated the setup. The diode parameters could be extracted directly from the HHCC data as it can be realized also on unknown samples.

\newpage

\section{Additional electrical transport data of the lithium niobate domain wall samples DW-1 and DW-2: Additional plots and full DC I-V curve fit parameters}

\begin{figure}[htb]
    \centering
    \includegraphics[width=\textwidth]{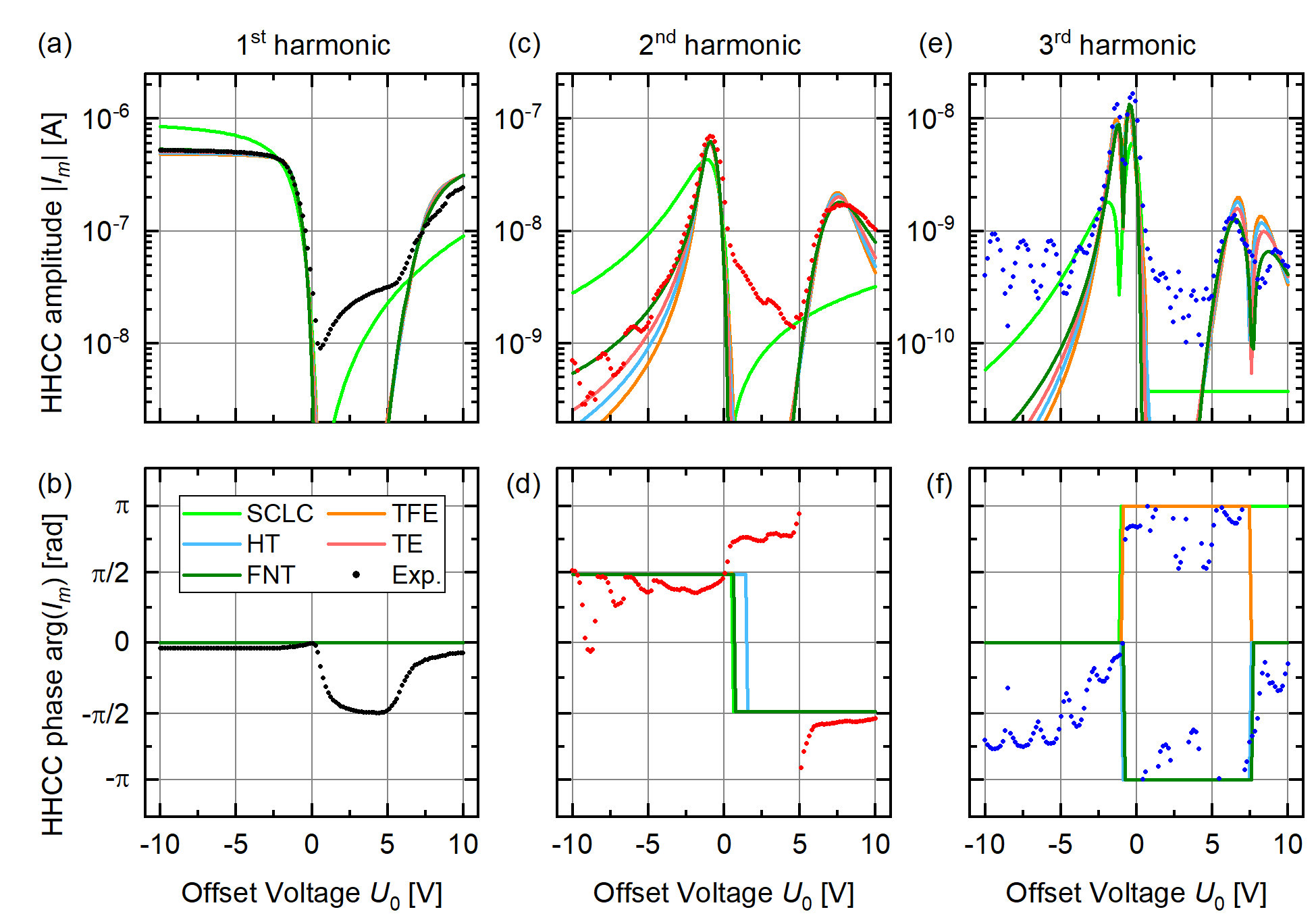}
    \vspace{-3mm}
    \caption{Complete measured and predicted HHCC data sets, i.e., amplitude \emph{and} phase, of sample \emph{DW-1} -- complementing fig.~5 of the main text. Concerning the \textit{X}-part of the \emph{R2X2} models, here also the discarded models of SCLC and TFE are included.}
    \label{fig:app:ac-models:dw1-full}
\end{figure}

\begin{figure*}[htb]
    \centering
    \includegraphics[width=\textwidth]{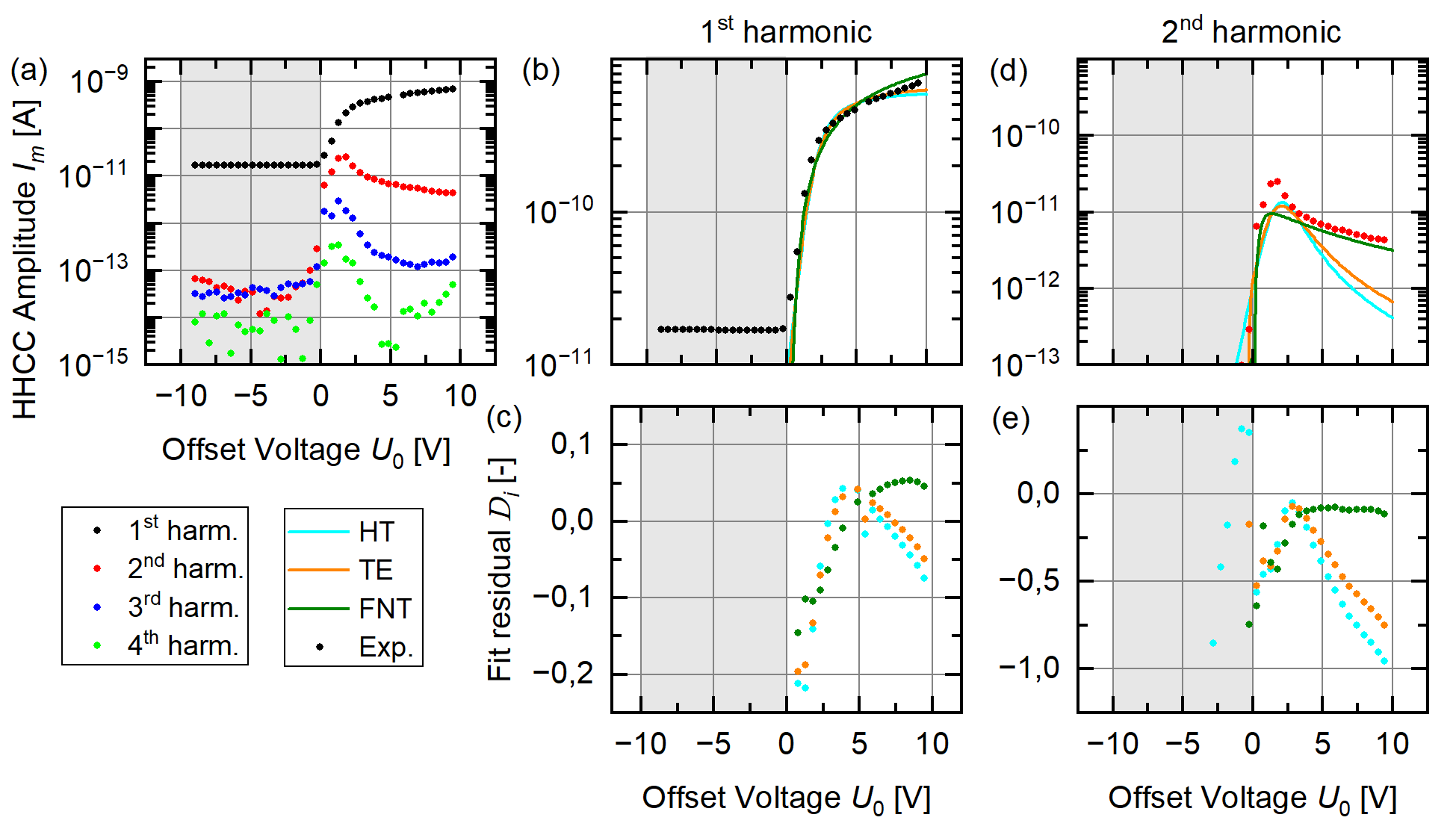}
    \caption{HHCC amplitudes as a function of the offset voltage $U_0$ at constant excitation amplitude ($U_1 = \SI{0.4}{V}$) and frequency ($\omega_1/2 \pi= \SI{38.5}{Hz}$) of sample \textit{DW-2}: Panel \textbf{(a)} comparatively shows the measured HHCC amplitudes for the first four harmonics, while panel \textbf{(b)} depicts the first-harmonic's experimental data \textit{together} with the predictions from the \emph{RX} equivalent circuit model with the \textit{X}-part being represented by the HT-, TE-, FNT-model, respectively. The corresponding residuals are plotted in panel \textbf{(d)}, while panels \textbf{(c)} and \textbf{(e)} show the analogous results for the case of the second-harmonic HHCC amplitudes. For the third- and fourth-harmonic cases, the results are not plotted, since the current values are around or below the detection limit.}
    \label{fig:app:dw2-hhcc-offset-predictions}
\end{figure*}

\begin{table}[htb]
    \centering
    {\renewcommand{\arraystretch}{1.3}
    \begin{tabular}{lrlrlrr} \hline \hline
        Model & \multicolumn{4}{c}{Parameters} & Residual $\mathcal{D}$ & $R^2$ \\ \hline
        \multirowcell{3}[0pt][l]{Hopping transport} 
            & $R_\text{f}$ & $= \SI{4.07 \pm 0.15}{M\ohm}$ & $R_\text{b}$ & $= \SI{3.12 \pm 0.22}{M\ohm}$ & \multirow{3}{*}{\num{1.50e-2}} & \multirow{3}{*}{\num{0.999307}} \\
            & $I_\text{HT, f}$ & $= \SI{6.3 \pm 5.1}{fA}$ & $I_\text{HT, b}$ & $= \SI{0.14 \pm 0.24}{nA}$ \\
            & $U_\text{HT, f}$ & $= \SI{452 \pm 24}{mV}$ & $U_\text{HT, b}$ & $= \SI{152 \pm 44}{mV}$ \\ \hline
        \multirow{3}{*}{Thermionic Emission} 
            & $R_\text{f}$ & $= \SI{3.83 \pm 0.1}{M\ohm}$ & $R_\text{b}$ & $= \SI{3.00 \pm 0.12}{M\ohm}$ & \multirow{3}{*}{\num{1.42e-2}} & \multirow{3}{*}{\num{0.999345}} \\
            & $I_\text{TE, f}$ & $= \SI{10.0 \pm 1.4e-21}{A}$ & $I_\text{TE, b}$ & $= \SI{0.76 \pm 0.63}{pA}$ \\
            & $U_\text{TE, f}$ & $= \SI{8.42 \pm 0.09}{mV}$ & $U_\text{TE, b}$ & $= \SI{7.3 \pm 0.1}{mV}$ \\ \hline
        \multirow{3}{*}{Thermionic Field Emission} 
            & $R_\text{f}$ & $= \SI{4.23 \pm 0.13}{M\ohm}$ & $R_\text{b}$ & $= \SI{3.22 \pm 0.14}{M\ohm}$ & \multirow{3}{*}{\num{1.69e-2}} & \multirow{3}{*}{\num{0.999158}} \\
            & $\sigma_\text{TFE, f}$ & $= \SI{3.1 \pm 0.9}{pS}$ & $\sigma_\text{TFE, b}$ & $= \SI{4.1 \pm 1.4}{nS}$ \\
            & $U_\text{TFE, f}$ & $= \SI{6.68 \pm 0.28}{V}$ & $U_\text{TFE, b}$ & $= \SI{551 \pm 45}{mV}$ \\ \hline
        \multirow{3}{*}{Fowler-Nordheim Tunneling} 
            & $R_\text{f}$ & $= \SI{3.19 \pm 0.1}{M\ohm}$ & $R_\text{b}$ & $= \SI{2.7 \pm 0.4}{M\ohm}$ & \multirow{3}{*}{\num{1.27e-2}} & \multirow{3}{*}{\num{0.999414}} \\
            & $\alpha_\text{FNT, f}$ & $= \SI{56 \pm 8}{\micro S/V}$ & $\alpha_\text{FNT, b}$ & $= \SI{2.1 \pm 2.8}{\micro S/V}$ \\
            & $U_\text{FNT, f}$ & $= \SI{79.4 \pm 0.1}{V}$ & $U_\text{FNT, b}$ & $= \SI{3.2 \pm 1.1}{V}$ \\ \hline
        \multirowcell{2}[0pt][l]{Space-charge \\ limited conduction} 
            & $R_\text{f}$ & $= \SI{0.037 \pm 508}{\kilo\ohm}$ & $R_\text{b}$ & $= \SI{1.5 \pm 0.6}{M\ohm}$ & \multirow{2}{*}{\num{2.82}} & \multirow{2}{*}{\num{0.869684}} \\
            & $\alpha_\text{SCLC, f}$ & $= \SI{1.9 \pm 0.2e-10}{\siemens/V}$ & $\alpha_\text{SCLC, b}$ & $= \SI{4.9 \pm 1.5e-8}{\siemens/V}$ \\ \hline \hline
    \end{tabular}}
    \caption{DC I-V curve analysis of \textit{DW-1}: fit parameters and residuals for different \textit{R2X2} models (cf. \cref{fig:ac_models:iucurves:DW1-current}), with the FNT model showing the lowest sum of residuals. However, the fact that the residuals and $R^2$ values for all considered models except the SCLC model are very close together, motivates to use an alternating-voltage excitation scheme and to analyze the resulting higher-harmonic current response in order to consolidate and verify the finding from the DC I-V curve fitting.}
    \label{tab:app:mechanisms_dc_parameter_dw1}
\end{table}

\begin{table}[htb]
    \centering
    {\renewcommand{\arraystretch}{1.3}
    \begin{tabular}{lrlrr} \hline \hline
        Model 
        & \multicolumn{2}{c}{Parameters} & Residual $\mathcal{D}$ & $R^2$ \\ \hline
        \multirowcell{3}[0pt][l]{Hopping transport} 
            & $R_\text{f}$  & $= \SI{991 \pm 27}{M\ohm}$ & \multirowcell{3}[0pt][l]{\num{7.38e-3}} & \multirowcell{3}[0pt][l]{\num{0.999673}} \\
            & $I_\text{HT, f}$ & $= \SI{5.9 \pm 0.7}{pA}$ \\
            & $U_\text{HT, f}$ & $= \SI{492 \pm 19}{mV}$ \\ \hline
        \multirowcell{3}[0pt][l]{Thermionic Emission} & 
            $R_\text{f}$ & $= \SI{889 \pm 22}{M\ohm}$  &  \multirowcell{3}[0pt][l]{\num{4.19e-3}} &  \multirowcell{3}[0pt][l]{\num{0.999815}} \\
            & $I_\text{TE, f}$ & $= \SI{208 \pm 29}{fS}$ \\
            & $U_\text{TE, f}$ & $= \SI{37.0 \pm 1.6}{mV}$ \\ \hline
         \multirowcell{3}[0pt][l]{Thermionic Field Emission} 
            & \(R_\text{f}\) & \(= \SI{1.091 \pm 0.032}{G\ohm}\) &  \multirowcell{3}[0pt][l]{\num{1.30e-2}} &  \multirowcell{3}[0pt][l]{\num{0.999423}} \\
            & \(\sigma_\text{TFE, f}\) & \(= \SI{24.7 \pm 2.1}{pS}\) \\
            & \(U_\text{TFE, f}\) & \(= \SI{2.00 \pm 0.15}{V}\) \\ \hline
         \multirowcell{3}[0pt][l]{Fowler-Nordheim Tunneling} 
            & \(R_\text{f}\) & \(= \SI{289 \pm 20}{M\ohm}\)  &  \multirowcell{3}[0pt][l]{\num{1.45e-3}} &  \multirowcell{3}[0pt][l]{\num{0.999936}} \\
            & \(\alpha_\text{FNT, f}\) & \(= \SI{1.40 \pm 5}{pS/V}\) \\
            & \(U_\text{FNT, f}\) & \(= \SI{1.46 \pm 0.46}{V}\) \\ \hline
         \multirowcell{2}[0pt][l]{Space-charge limited conduction} 
            & \(R_\text{f}\) & \(= \SI{587 \pm 8}{M\ohm}\) &  \multirowcell{2}[0pt][l]{\num{1.55e-3}} &  \multirowcell{2}[0pt][l]{\num{0.999931}} \\
            & \(\alpha_\text{SCLC, f}\) & \(= \SI{36.0 \pm 0.4}{p\siemens/V}\) \\ \hline \hline
    \end{tabular}}
    \caption{DC I-V curve analysis of \textit{DW-2}: fit parameters and residuals for different \textit{RX} models (cf. \cref{fig:ac_models:iucurves:DW2-current}), with the FNT model showing the lowest sum of residuals. However, the fact that the residuals and $R^2$ values for all considered models except the SCLC model are very close together, motivates us to use an alternating-voltage excitation scheme and to analyze the resulting higher-harmonic current response in order to consolidate and verify the finding from the DC I-V curve fitting.}
    \label{tab:app:mechanisms_dc_parameter_dw2}
\end{table}

\end{document}